\documentclass[a4paper,twocolumn,11pt, accepted=2024-01-02]{quantumarticle}
\pdfoutput=1
\usepackage[utf8]{inputenc}
\usepackage[english]{babel}
\usepackage[T1]{fontenc}
\usepackage{amsthm}
\usepackage{amssymb}
\usepackage{amsmath}
\usepackage{mathtools}
\usepackage[ruled,linesnumbered]{algorithm2e}
\usepackage{hyperref}
\usepackage{xspace}
\usepackage{braket}
\usepackage[numbers]{natbib}
\usepackage{url}
\usepackage{systeme}
\usepackage{appendix}
\hypersetup{pdfpagemode=UseNone}
\usepackage{physics}
\usepackage{xcolor}
\usepackage{siunitx}
\usepackage{multirow}
\usepackage{booktabs}

\newcommand{\inner}[2]{\langle #1|#2 \rangle}

\begin{document}

\title{A full circuit-based quantum algorithm for excited-states in quantum chemistry}

\author{Jingwei Wen}
\affiliation{State Key Laboratory of Low-Dimensional Quantum Physics and Department of Physics, Tsinghua University, Beijing 100084, China}
\affiliation{China Mobile (Suzhou) Software Technology Company Limited, Suzhou 215163, China}
\email{wjw17@tsinghua.org.cn}

\author{Zhengan Wang}
\affiliation{Beijing Academy of Quantum Information Sciences, Beijing 100193, China}

\author{Chitong Chen}
\affiliation{Institude of Physics, Chinese Academy of Sciences, Beijing 100190, China}
\affiliation{School of Physical Sciences, University of Chinese Academy of Sciences, Beijing 100190, China}

\author{Junxiang Xiao}
\affiliation{State Key Laboratory of Low-Dimensional Quantum Physics and Department of Physics, Tsinghua University, Beijing 100084, China}

\author{Hang Li}
\affiliation{Beijing Academy of Quantum Information Sciences, Beijing 100193, China}

\author{Ling Qian}
\affiliation{China Mobile (Suzhou) Software Technology Company Limited, Suzhou 215163, China}

\author{Zhiguo Huang}
\affiliation{China Mobile (Suzhou) Software Technology Company Limited, Suzhou 215163, China}

\author{Heng Fan}
\affiliation{Beijing Academy of Quantum Information Sciences, Beijing 100193, China}
\affiliation{Institude of Physics, Chinese Academy of Sciences, Beijing 100190, China}

\author{Shijie Wei}
\affiliation{Beijing Academy of Quantum Information Sciences, Beijing 100193, China}
\email{weisj@baqis.ac.cn}

\author{Guilu Long}
\email{gllong@tsinghua.edu.cn}
\affiliation{State Key Laboratory of Low-Dimensional Quantum Physics and Department of Physics, Tsinghua University, Beijing 100084, China}
\affiliation{Beijing Academy of Quantum Information Sciences, Beijing 100193, China}
\affiliation{Frontier Science Center for Quantum Information, Beijing 100084, China}
\affiliation{Beijing National Research Center for Information Science and Technology, Beijing 100084, China}

\maketitle


\begin{abstract}

Utilizing quantum computer to investigate quantum chemistry is an important research field nowadays. In addition to the ground-state problems that have been widely studied, the determination of excited-states plays a crucial role in the prediction and modeling of chemical reactions and other physical processes. Here, we propose a non-variational full circuit-based quantum algorithm for obtaining the excited-state spectrum of a quantum chemistry Hamiltonian. Compared with previous classical-quantum hybrid variational algorithms, our method eliminates the classical optimization process, reduces the resource cost caused by the interaction between different systems, and achieves faster convergence rate and stronger robustness against noise without barren plateau. The parameter updating for determining the next energy-level is naturally dependent on the energy measurement outputs of the previous energy-level and can be realized by only modifying the state preparation process of ancillary system, introducing little additional resource overhead. Numerical simulations of the algorithm with hydrogen, LiH, H2O and NH3 molecules are presented. Furthermore, we offer an experimental demonstration of the algorithm on a  superconducting quantum computing platform, and the results show a good agreement with theoretical expectations. The algorithm can be widely applied to various Hamiltonian spectrum determination problems on the fault-tolerant quantum computers.

\end{abstract}


\section{Introduction}

As one of the emerging research fields, quantum computation is devoted to solving some certain computational problems that are intractable to deal with in classical computers using the principles of quantum mechanics \cite{benioff1980computer,feynman2018simulating}. Since the concept was proposed, quantum computing has become one of the most fruitful fields in contemporary physics and various important problems of practical significance such as prime factorization \cite{shor1999polynomial}, database search \cite{grover1997quantum,long1999phase}, and solution of linear equations \cite{harrow2009quantum,subacsi2019quantum} have been solved by algorithms in quantum version. In recent years, the development of quantum algorithms applied to quantum chemistry have become an active research field, with huge potential market application value \cite{cao2019quantum,mcardle2020quantum,bauer2020quantum}. Some quantum algorithms used to determine the ground-state of chemical molecule Hamiltonian, such as the classical-quantum hybrid variational quantum eigensolver (VQE) algorithm \cite{peruzzo2014variational,o2016scalable,kandala2017hardware,cerezo2021variational} and its improvements \cite{bonet2018low,grimsley2019adaptive,tang2021qubit,ostaszewski2021structure}, the full quantum eigensolver (FQE) algorithm \cite{wei2020full}, have obtained rapid theoretical development and some state-of-the-art experimental demonstration. To be specific, the VQE algorithm is divided into two parts, i.e. preparing and measuring of quantum states on the quantum computer and parameter optimization on the classical computer. By iterating the whole process until convergence, the ground-state and ground-state energy of the target Hamiltonian can be obtained. But the FQE algorithm removes the classical optimizer and performs all the calculations on the quantum computer by using quantum gradient descent \cite{rebentrost2019quantum}.

However, in addition to the ground-states, the determination of the excited-states is also indispensable for the study of chemical reaction processes. Up to now, a series of modified versions of the hybrid variational algorithms have been proposed for finding energy spectrum, such as variational quantum deflation (VQD) \cite{higgott2019variational, jones2019variational} and subspace-search variational quantum eigensolver (SSVQE) \cite{nakanishi2019subspace,parrish2019quantum} algorithms and some extension versions \cite{mcclean2017hybrid,colless2018computation,jouzdani2019method,ollitrault2020quantum,zhang2020variational,yalouz2021analytical,wen2021variational}. The basic idea of the former method is introducing state-specific penalization terms to the Hamiltonian and determining each eigenstate by a separate minimization. While the latter algorithm performs a single minimization for a set of initially selected orthogonal states with one ansatz, and realizes the mapping from these states to the lower excited-states of target Hamiltonian.

\begin{figure}
\centering
\includegraphics[width=\linewidth]{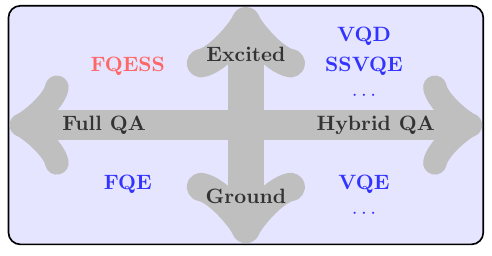}
\caption{Classification of quantum algorithms for quantum chemistry. The algorithms are divided into hybrid quantum algorithm for noisy intermediate scale quantum (NISQ) devices and full quantum algorithm for fault-tolerant quantum computers, where the former one requires the cooperation of classical and quantum computers while the latter one is based on the full circuit-based quantum operation. The purpose of quantum chemistry is divided into determining ground- or excited-states.}
\label{fig_fourXX} 
\end{figure}

Although so much progress has been made in the investigation of excited-states, it is still a research direction of concern to determine the energy spectrum of the molecular Hamiltonian based on a complete quantum circuit model for future fault-tolerant quantum computation, just as shown in the Figure \ref{fig_fourXX}. In this work, we fill the last step of solving quantum chemistry problems in different algorithm frames and propose a full quantum excited-state solver (FQESS) algorithm for determining the whole spectrum of chemistry Hamiltonian efficiently and steadily. Compared with classical-quantum hybrid variational algorithms, our method removes the optimization in classical computers, and its non-variational nature can ensure that the algorithm converges to the target states along the direction of the fastest gradient descent, avoiding barren plateau phenomenon. Moreover, the parameter updating for different energy-levels can be simply realized by modifying the state preparation process of ancillary system based on the energy measurement of the last energy-level, which is experimentally friendly. 

This paper is organized as follows: in Sec \ref{Chaptheory}, we introduce the details of the FQESS algorithm, and analyze the complexity of the algorithm. In Sec \ref{Chapsimulation}, we present a numerical simulation with hydrogen, LiH, H2O and NH3 molecules under noiseless and noise condition separately. In Sec \ref{Chapexp}, we offer an experimental demonstration of FQESS algorithm on the real superconducting quantum computing platform. Finally, sec \ref{Chapconclusion} gives a conclusion.


\section{Results} \label{Chaptheory}

\subsection{Full Quantum Excited-State Solver}

Here we introduce a full quantum algorithm for solving the excited-states of the quantum chemistry Hamiltonian. Quantum simulation of fermionic systems can be reformulated in terms of qubit operations by Jordan-Wigner \cite{jordan1993paulische} or Bravyi-Kitaev transformation \cite{bravyi2002fermionic}, and then the target $n$-qubit chemistry Hamiltonian can be generally expressed as
\begin{equation}
H_{1}=\sum_{i=1}^{L_{1}}\alpha_{i}^{(1)}P_{i}
\label{H1}
\end{equation}
which is expressed as the linear combination of $L_{1}\le4^{n}$ Pauli words $P_{i}$ (tensor product of Pauli matrices) with real coefficients $\alpha_{i}^{(1)}$, and we want to find the spectrum. Let the set of eigenstates be $\{\ket{E_{j}}\}_{j=1}^{2^{n}}$ with corresponding eigenenergies $\{E_{j}\}_{j=1}^{2^{n}}$, satisfying $\vert E_{i} \vert \ge \vert E_{j} \vert $ when $i\le j$. 

\begin{figure}
\centering
\includegraphics[width=\linewidth]{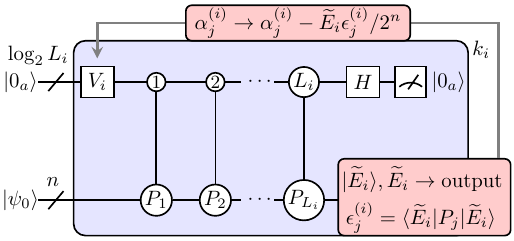}
\caption{Quantum circuit for the realization of FQESS algorithm.  The whole system includes $n+\log_{2}L_{i}$ qubits and is divided into working system and ancillary system. The basic process for determining the $i$-th eigenstate, which is repeated by $k_{i}$ times, includes four parts, that is encoding with operator $V_{i}$, entangling with $L_{i}$ controlled gates, decoding with Hardmard gates and measurement on the ancillary qubits. The output state of working system is $\ket{\widetilde{E}_{i}}$ and its measurement results on the different Pauli words $\epsilon_{j}^{(i)}$ is used for calculating the eigenvalues $\widetilde{E}_{i}$ and updating the state preparation operator. }
\label{fig_circuit} 
\end{figure} 

We first need to construct a quantum circuit to realize the Hamiltonian in equation (\ref{H1}), as shown in the Figure \ref{fig_circuit}. Note that we have $n$-qubit as the working system, while we need another $\log_{2}L_{1}\le 2n$ qubits as the ancillary system. In principle, we at most need $3n$ qubits to ensure the expansion generality of the subsequent quantum states. The details for the realization of such a process with quantum circuits will be presented in section \ref{circuitU}. Once we realize the Hamiltonian $H_{1}$ with the quantum circuit, we apply the circuit-based operators with enough times ($k_{1}$ times) to an arbitrary initial quantum state $\ket{\psi_{0}}$, then we have the normalized quantum state as $\ket{\widetilde{E}_{1}}=H^{k_{1}}_{1}\ket{\psi_{0}}/(\bra{\psi_{0}}H^{2k_{1}}_{1}\ket{\psi_{0}})^{1/2}$, an approximation to the eigenvector $\ket{E_{1}}$ with the biggest absolute value of eigenvalue $\vert E_{1} \vert$. The corresponding approximate eigenvalue $\widetilde{E}_{1}$ can be determined by $\widetilde{E}_{1}=\bra{\widetilde{E}_{1}}H_{1}\ket{\widetilde{E}_{1}}=\sum_{i=1}^{L_{1}}\alpha_{i}^{(1)}\epsilon_{i}^{(1)}$, and the energy component $\epsilon_{i}^{(1)}=\bra{\widetilde{E}_{1}}P_{i}\ket{\widetilde{E}_{1}}$ can be obtained by measuring the average values of the output quantum state under different Pauli words. 

Furthermore, based on the completeness of Pauli basis, output density matrix $\rho_{1}=\ket{\widetilde{E}_{1}}\bra{\widetilde{E}_{1}}$ can be expanded as $\rho_{1}=\sum_{j=1}^{l_{1}}\beta_{j}^{(1)}P_{j}$, where the coefficients satisfy $\beta_{j}^{(1)}=\textup{tr}(P_{j}\rho_{1})/2^{n}=\epsilon_{j}^{(1)}/2^{n}$. This can be understood as the energy contribution of $\rho_{1}$ to each Pauli term in the Hamiltonian. Then we can re-constructed the original Hamiltonian as 
\begin{equation}
H_{2}=\sum_{i=1}^{L_{2}}(\alpha_{i}^{(1)}-\widetilde{E}_{1}\beta_{i}^{(1)})P_{i}=\sum_{i=1}^{L_{2}}\alpha_{i}^{(2)}P_{i}
\end{equation}
which can be realized with the same circuit as above, just changing the initial state preparation of ancillary system. The corresponding physical meaning is that we have eliminated the energy contribution of $\ket{E_{1}}$ to each Pauli term in the Hamiltonian, so the biggest absolute value of eigenvalues for the new Hamiltonian is $\vert E_{2} \vert$ now. Similarly, by applying the new circuit-based operator with $k_{2}$ times to the arbitrary initial quantum state, we can have 
\begin{equation}
\ket{\widetilde{E}_{2}}=\frac{H^{k_{2}}_{2}\ket{\psi_{0}}}{\sqrt{\bra{\psi_{0}}H^{2k_{2}}_{2}\ket{\psi_{0}}}},~~
\widetilde{E}_{2}=\sum_{i=1}^{L_{2}}\alpha_{i}^{(2)}\epsilon_{i}^{(2)}
\end{equation}
which is the eigenvector $\ket{E_{2}}$ with the second biggest absolute value of eigenvalue in $H_{1}$ and corresponding eigenvalue $E_{2}$.

Repeat this process with $2^{n}$ times, we can determine the whole spectrum of the initial Hamiltonian. Note that the order of solution is based on the magnitude of the absolute values of the eigenvalues, rather than the magnitude of the eigenvalues themselves, which may cause some confusion. This can be solved by introducing a bias term $-\lambda_{0}I^{\otimes n}~(\lambda_{0}>\textup{max}\{0, E_{1}, \cdots E_{2^{n}}\})$ into the Hamiltonian, making all the eigenvalues negative. Then the first solved eigenstate is the ground-state, followed by the first excited-state, and so on. And the bias parameter $\lambda_{0}$ should not be too big, because it affects the ratio of reconstructed eigenvalues $(E_{i}-\lambda_{0})/(E_{1}-\lambda_{0})$, which is related to the convergence rate of the algorithm. In general, the bigger the bias parameter is, the more operation times we need. The repetition satisfies $k_{i}=\mathcal{O}(\log(N/\epsilon))$, which is logarithmically dependent on the system size $N$ and the inverse of energy precision $\epsilon$ \cite{wei2020full}. The FQESS algorithm proposed here can be integrated into the scope of power iteration scheme or quantum gradient descent scheme for eigenvalue evaluation, as discussed in the appendix \ref{FQESSbaisc}. The detailed process of the FQESS algorithm is summarized in the table of Algorithm \ref{AlgorithmProcess}.


\begin{algorithm}[t]
\caption{FQESS algorithm}
\label{AlgorithmProcess}
\KwIn{Hamiltonian $H_{1}=\sum_{j=1}^{L_{1}\le4^{n}}\alpha_{j}^{(1)}P_{j}$, State $\ket{\psi_{0}}$, Bias parameter $\lambda_{0}$, Iteration times $k_{i}$}
\KwOut{Eigenstates $\ket{\widetilde{E}_{i}}$, Eigenvalues $\widetilde{E}_{i}$ \\
{\bf Preprocess:} Construct $U_{1}=H_{1}-\lambda_{0}I^{\otimes n}$ } 
\For{ $i=1:2^{n}$}
{Apply circuit $k_{i}$ times to $\ket{\psi_{0}}$, having $\ket{\widetilde{E}_{i}}=U_{i}^{k_{i}}\ket{\psi_{0}}$ \;
Measure $\epsilon_{j}^{(i)}=\bra{\widetilde{E}_{i}}P_{j}\ket{\widetilde{E}_{i}}$ and get $\widetilde{E}_{i}=\sum_{j=1}^{L_{i}}\alpha_{j}^{(i)}\epsilon_{j}^{(i)}$ \;
{\bf return} $\ket{\widetilde{E}_{i}}$ and $\widetilde{E}_{i}$ \;
Reconstruct $\alpha_{j}^{(i+1)} = \alpha_{j}^{(i)}-\widetilde{E}_{i}\epsilon_{j}^{(i)}/2^{n}$ \;
Construct circuit $U_{i+1}=\sum_{j=1}^{L_{i+1}}\alpha_{j}^{(i+1)}P_{j}-\lambda_{0}I^{\otimes n}$
}
\end{algorithm}



\subsection{Realization of Circuit-Based Operator}\label{circuitU}

Now we turn to discuss concretely about how to realize the operator $U_{i}=H_{i}-\lambda_{0}I^{\otimes n}$ in the circuit-based quantum computation frame. Without loss of generality, we can set $P_{1}=I^{\otimes n}$, then we have
\begin{equation}
U_{i}=(\alpha_{1}^{(i)}-\lambda_{0})I^{\otimes n}+\sum_{j=2}^{L_{i}}\alpha_{j}^{(i)}P_{j}
\label{Ui}
\end{equation}
where $L_{i}\le 4^{n}$. This is the target operator we aimed to repeat by $k_{i}$ times when determining the $i$-th eigenstate. In principle, it can be understood as a linear combination of unitary Pauli words and can be simulated by introducing ancillary qubits to form a bigger Hilbert space \cite{gui2006general,gui2008duality,gui2009allowable,childs2012hamiltonian,wen2019experimental,wen2020observation}, as shown in the Figure \ref{fig_circuit}. The basic process is divided into four parts, including encoding, entangling, decoding and measurement. The first encoding process is a quantum state preparation process for ancillary system, realized by the $\log_{2}L_{i}$-qubit operator $V_{i}$, whose first column is $ [ \alpha_{1}^{(i)}-\lambda_{0},\alpha_{2}^{(i)},\cdots,\alpha_{L_{i}}^{(i)} ]$. It does not matter what the other matrix elements are as long as the operator is unitary, and we can determine the operator by schmidt orthogonalization or by decomposing it into single- and two-qubit operators. Selectively, the quantum state of ancillary system after encoding is 
\begin{equation}
V_{i}\ket{0_{a}}=\Big[(\alpha_{1}^{(i)}-\lambda_{0})\ket{1}+\sum_{j=2}^{L_{i}}\alpha_{j}^{(i)}\ket{j}\Big]/\mathcal{C}
\end{equation}
where $\mathcal{C}=\big[(\alpha_{1}^{(i)}-\lambda_{0})^2+\sum_{j=2}^{L_{i}}(\alpha_{j}^{(i)})^{2}\big]^{1/2}$ is a normalization constant. Note that the notations $\ket{0_{a}}$ and $\ket{1}$ refer to the same quantum states $\ket{0^{\otimes  \log_{2}L_{i}}}$ here. This quantum state can be prepared by the initialization algorithm in Ref \cite{long2001efficient} with $\mathcal{O}(L_{i}\log_{2}L_{i})$ standard gate operations or $\mathcal{O}(\log_{2}L_{i})$ steps with quantum random access memory method \cite{giovannetti2008quantum}. Then a series of multi-qubit controlled operators $\sum_{i=1}^{L_{i}}\ket{i}\bra{i}P_{i}$ are applied onto the whole system, creating entanglement between the ancillary and working systems. The state of the whole system is transformed into $\big [\ket{1}(\alpha_{1}^{(i)}-\lambda_{0})P_{1}+\sum_{j=2}^{L_{i}}\ket{j}\alpha_{j}^{(i)}P_{j} \big ]\ket{\psi_{0}}/\mathcal{C}$. Followed by the decoding operation realized with the Hardmard gates $H^{\otimes \log_{2}L_{i}}$ on the ancillary system, we concern the output state of working system in the $\ket{0^{\otimes \log_{2}L_{i}}}$ subspace of the ancillary system as
\begin{equation}
\frac{(\alpha_{1}^{(i)}-\lambda_{0})P_{1}+\sum_{j=2}^{L_{i}}\alpha_{j}^{(i)}P_{j}}{\mathcal{C}\sqrt{2^{L_{i}}}}\ket{\psi_{0}}=\frac{U_{i}\ket{\psi_{0}}}{\mathcal{C}\sqrt{2^{L_{i}}}}
\end{equation}
and this means that we can realize the simulation of target operator by measuring the ancillary system in state $\ket{0^{\otimes \log_{2}L_{i}}}$ with success probability $P_{s}^{(i)}=\vert\vert U_{i}\ket{\psi_{0}} \vert\vert^{2}/(\mathcal{C}^{2}2^{L_{i}})$ at each iteration, where $\vert\vert x \vert\vert$ represents the modulus of vector $x$. We can also further increase the probability for obtaining target quantum states by quantum amplitude amplification technology \cite{brassard2002quantum,berry2015simulating}.

Then the output of this basic process will be the input for the next iteration, which needs to be repeated $k_{i}$ times when determining the $i$-th eigenstate, and the final output state $\ket{\widetilde{E}_{i}}$ of the working system would be a good approximation to the eigenstate $\ket{E_{i}}$. Based on the output state, we can obtain its measurement values for each Pauli words $\epsilon_{j}^{(i)}=\bra{\widetilde{E}_{i}}P_{j}\ket{\widetilde{E}_{i}}$ as the basic energy components, which are then used for calculating the approximate eigenvalues $\widetilde{E}_{i}$ and updating the state preparation operator for determining the next eigenstate. It is worth emphasizing that when we want to determine different energy-levels, only the operator $V_{i}$ needs to be changed, which reduces the complexity and resource cost during the iteration updating process. In addition to the programmable or updatable state preparation process, the other parts of our algorithm can be modularized  in principle for determining the energy spectrum of different molecules. Moreover, the procedures for determining the $i$-th eigenstate in our FQESS algorithm is similar to that in FQE algorithm, and by updating the state preparation process using the energy output of last iteration, we extend the solution of the ground-state to the entire energy spectrum.

\begin{figure}
\centering
\includegraphics[width=\linewidth]{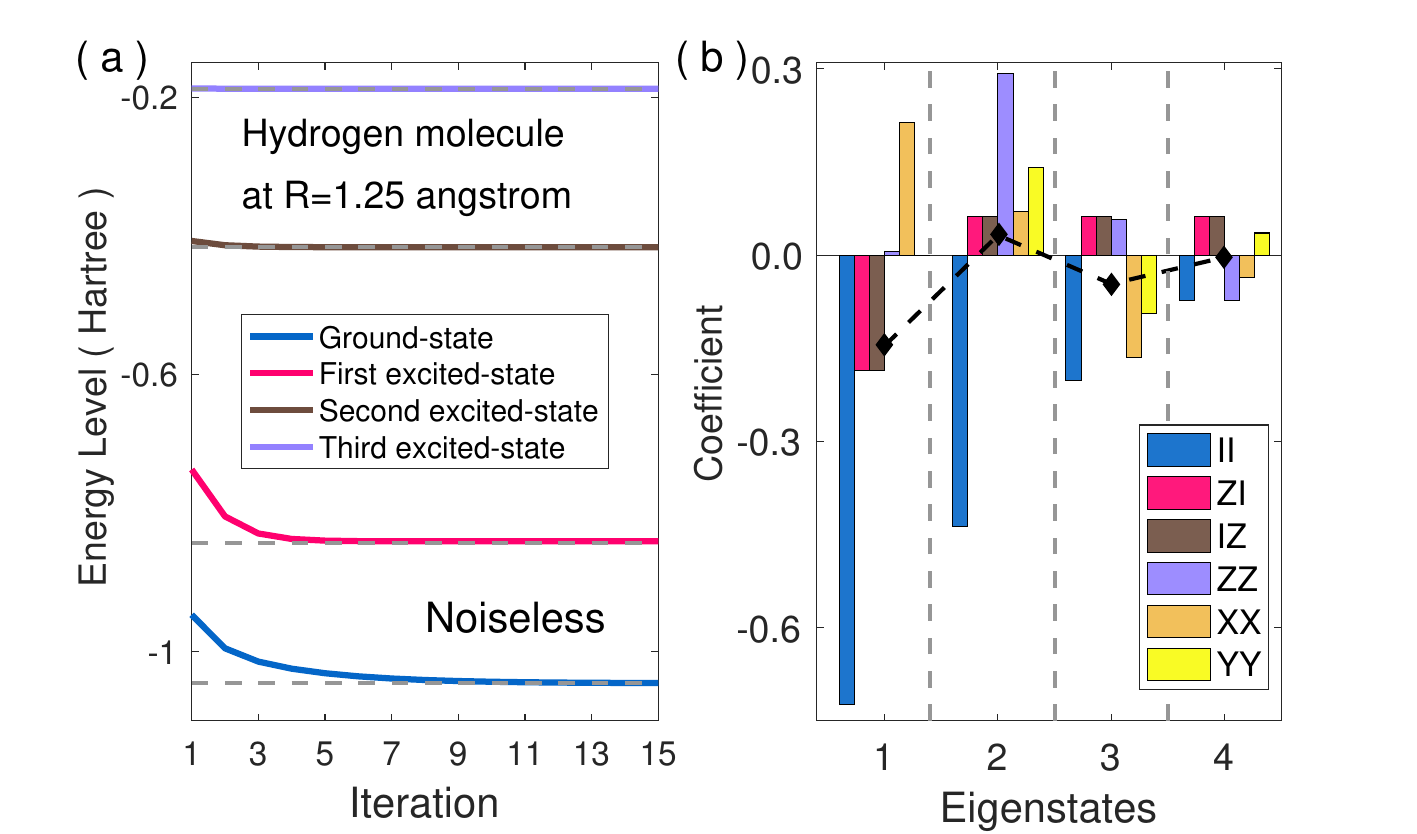}
\caption{A simulation of the iteration process for hydrogen molecule without noise at the $R=1.25$ angstrom. (a) The iteration processes for different energy-levels. (b) The updating parameter $\alpha_{j}^{(i)}$ under Pauli basis and the black diamond points represent the mean coefficients for different energy-levels. }
\label{fig_iteration_noiseless} 
\end{figure} 


\subsection{Complexity Analysis}

The algorithm complexity for our FQESS algorithm includes qubit resources and gate complexity. For qubit resources, the number of ancillary qubits needed is $\log_{2}L\le 2n$, where $L=\textup{max}\{L_{i}\}$ is the maximum number of Pauli words in the Hamiltonian. Therefore, the total number of qubits in our algorithm is $\mathcal{O}(n+\log_{2}L)$, no more than $3n$-qubits. As for the gate complexity in each basic process, we need $\mathcal{O}(L\log_{2}L)$ single- and two-qubit gates for the encoding process \cite{long2001efficient} and $\mathcal{O}(nL\log_{2}L)$ basic gates for entangling process \cite{xin2017quantum,wei2018efficient}. Consider another $\log_{2}L$ Hardmard gates for decoding, the total basic gates required for the realization of $U_{i}$ operator when obtaining the $i$-th energy-level is $\mathcal{O}(nL\log_{2}L)$ for the FQESS algorithm. For the chemistry Hamiltonian of electrons with $L=\mathcal{O}(n^{4})$, the gate complexity is $\mathcal{O}(n^{5}\log_{2}n)$. Moreover, to estimate the approximate eigenvalue $\widetilde{E}_{i}=\sum_{j=1}^{L_{i}}\alpha_{j}^{(i)}\epsilon_{j}^{(i)}$, we need $\mathcal{O}(L_i/P_{s}^{(i)}/\delta^2)$ measurements, where $P_{s}^{(i)}$ indicates the probability that all qubits of the auxiliary system are measured to be $\ket{0}$ quantum states, and $\delta\propto N^{-1/2}$ represents the statistical error of $\epsilon_{j}^{(i)}$ when $N$ measurements are made \cite{napolitano2011interaction}.


\begin{figure}
\centering
\includegraphics[width=\linewidth]{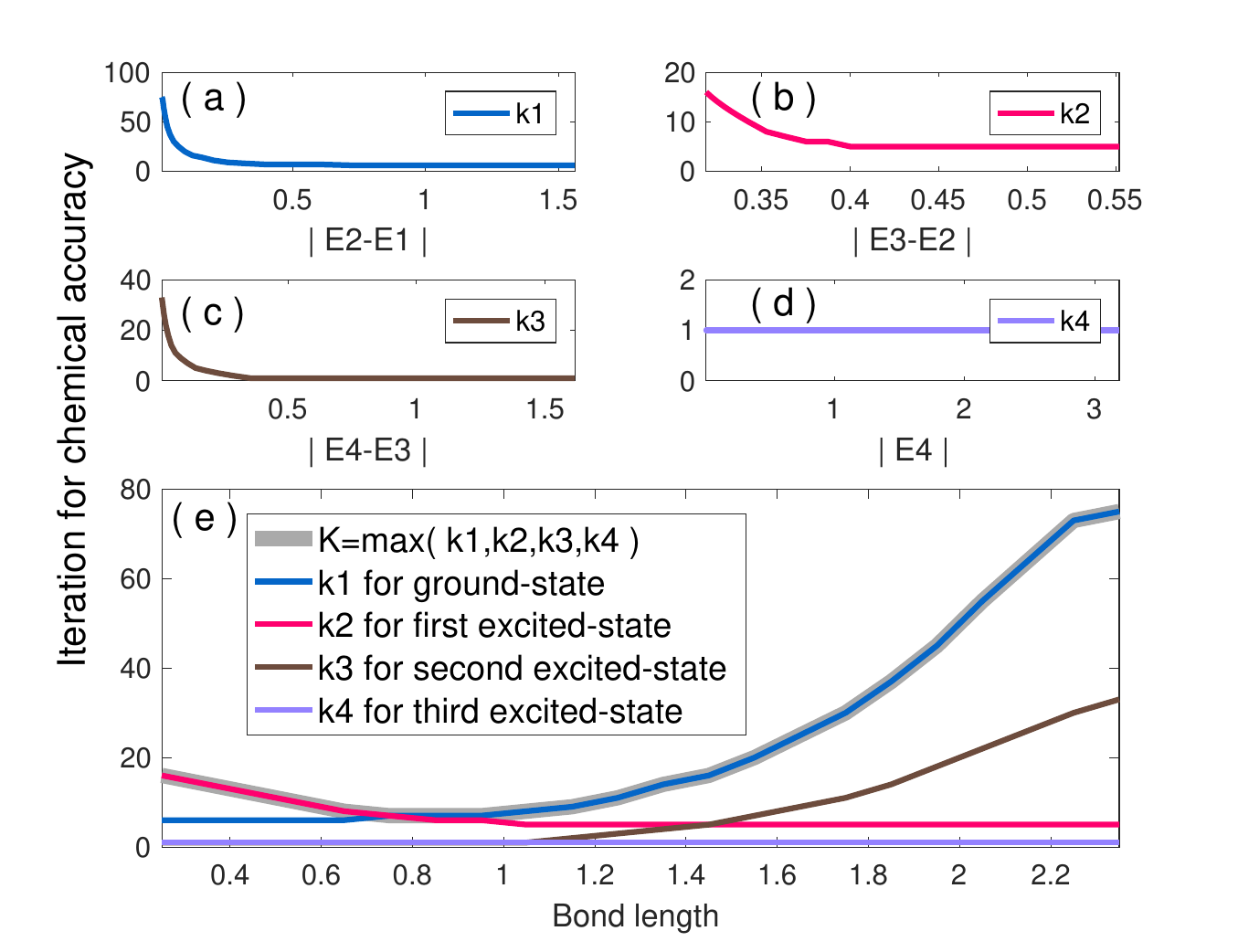}
\caption{(a-d) The minimum iteration times $k_{i}$ for realizing chemical accuracy when determining eigenenergy and its relation with the energy difference. (e) The maximum iteration times $K=\textup{max}\{k_{i}\}$ under different bond lengths (unit in angstrom).}
\label{fig_K} 
\end{figure}

\section{Numerical Simulation} \label{Chapsimulation}

\begin{figure*}
\centering
\includegraphics[width=0.49\linewidth]{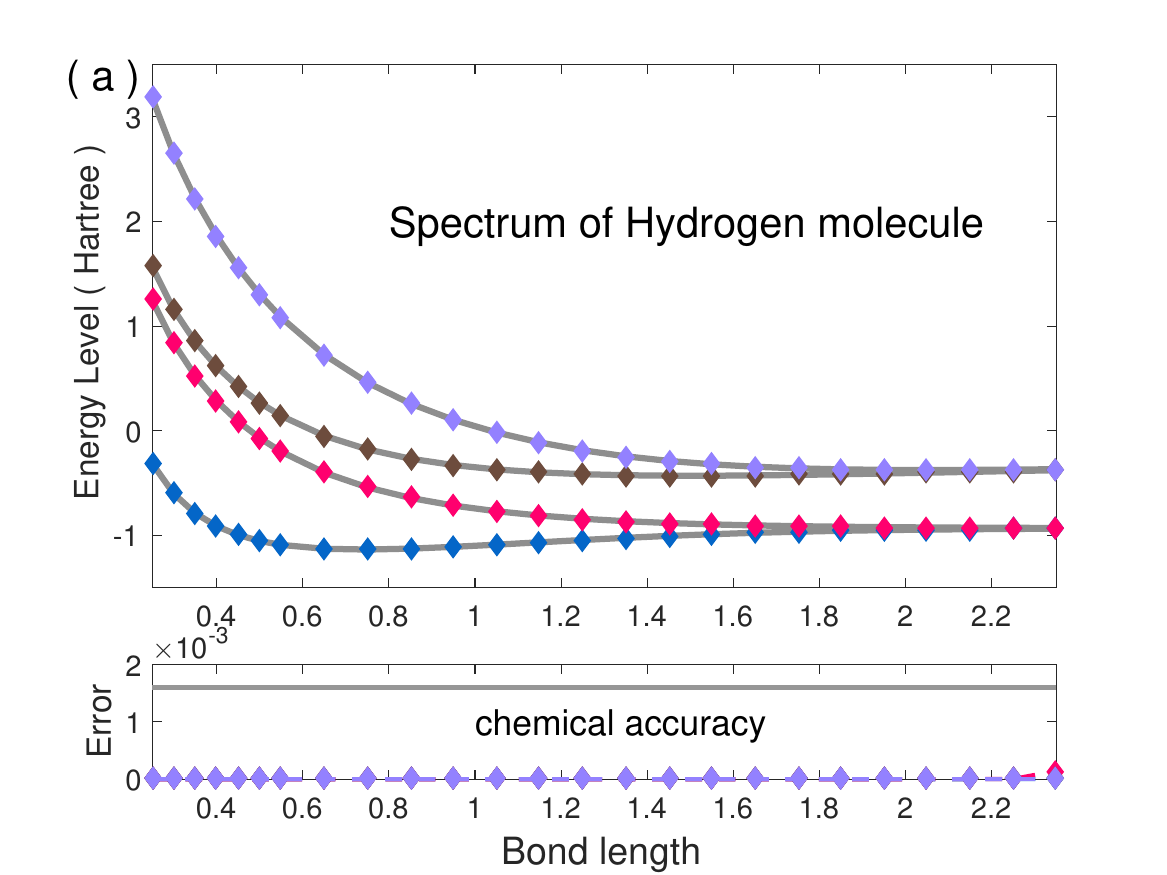}
\includegraphics[width=0.49\linewidth]{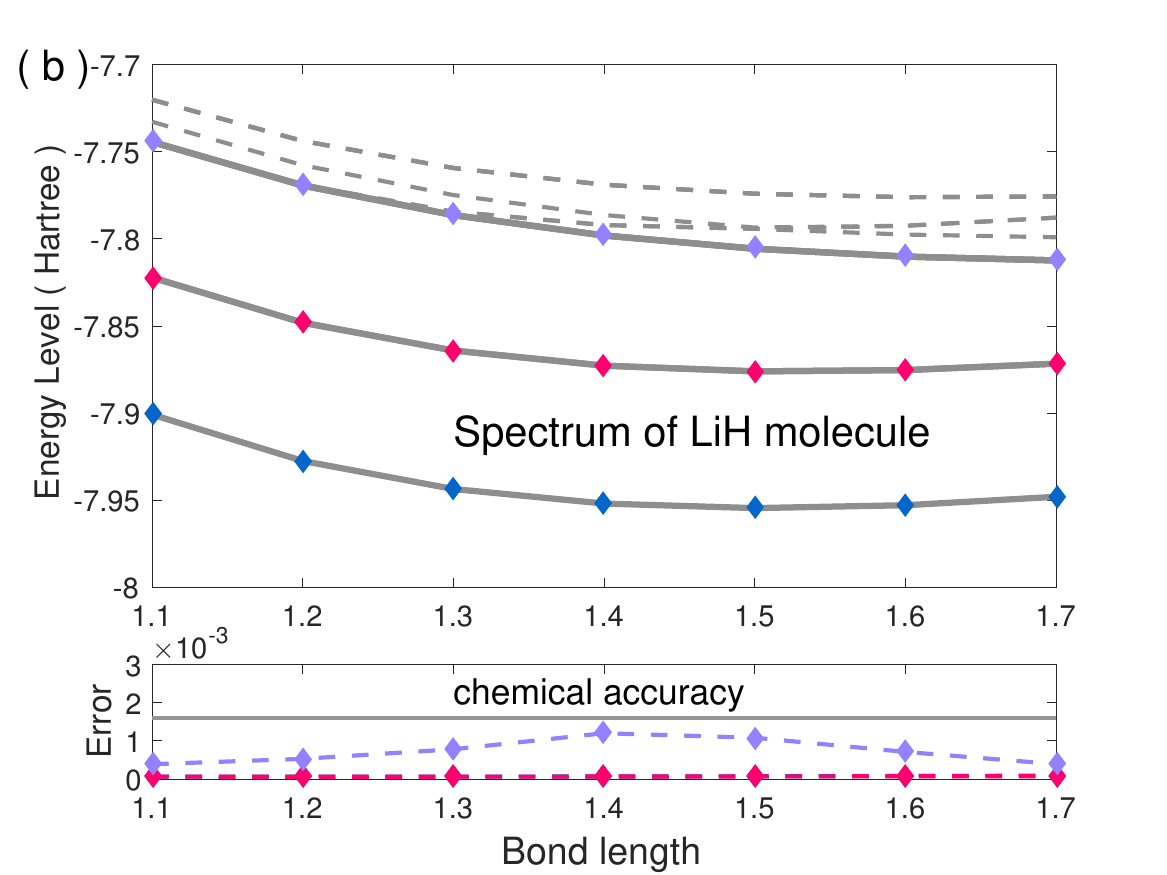}
\caption{A simulation of hydrogen (a) and LiH (b) molecule spectrum without noise for a range of inter-nuclear separations (unit in angstrom). The exact values are obtained by Hamiltonian diagonalization and plotted in gray lines. The numerical results are plotted with circles. Errors between the numerical outputs and the theoretical expectations are shown in the bottom panel.  }         
\label{fig_spectrum_noiseless} 
\end{figure*}

In this part, we present a demonstration of the FQESS algorithm for excited-states with the hydrogen and LiH molecules in the minimal STO-3G basis for a range of inter-nuclear separations to verify the feasibility and robustness of the algorithm. The fermionic Hamiltonian of hydrogen can be translated into qubit representation by Bravyi-Kitaev transformation \cite{bravyi2002fermionic}, obtaining a two-qubit Hamiltonian as
\begin{equation}
\begin{split}
H(R)=&\alpha_{0}^{R}+\alpha_{1}^{R}\sigma_{z}^{(1)}+\alpha_{2}^{R}\sigma_{z}^{(2)}+\alpha_{3}^{R}\sigma_{z}^{(1)}\otimes\sigma_{z}^{(2)} \\
&+\alpha_{4}^{R}\sigma_{x}^{(1)}\otimes\sigma_{x}^{(2)}+\alpha_{5}^{R}\sigma_{y}^{(1)}\otimes\sigma_{y}^{(2)}
\end{split}
\end{equation}
where $\sigma_{\beta}^{(i)}~(\beta=x,y,z)$ is the Pauli operator acting on the $i$-th qubit and the real-valued coefficients $\alpha_{i}^{R}$ are functions of the inter-nuclear distance $R$ \cite{colless2018computation}. Also, a six-qubit Hamiltonian for LiH molecule containing 118 Pauli words is obtained via the Jordan-Wigner transformation  \cite{jordan1993paulische}. Moreover, numerical simulation results of larger molecules for H2O (12 qubits) and NH3 (14 qubits) in STO-6G basis set are presented in Appendix \ref{appendixbigger}. First, we apply our FQESS algorithm to the hydrogen molecule and plot the power iteration process for each eigenstate at $R=1.25$ angstrom in the Figure \ref{fig_iteration_noiseless}, together with a updating parameter $\alpha_{j}^{(i)}$  under different Pauli basis. The initial state is chosen as $\ket{\psi_{0}}=\ket{0}\otimes\ket{+}$, where $\ket{+}$ is a eigenvector of $\sigma_{x}$ matrix. It can be concluded that iteration processes can quickly converge to the target values. More importantly, the iteration times needed is less when the energy-level is high enough, especially for the last eigenstate, there is no need for extra iterations. This is due to that our algorithm eliminates the other eigen-components during the iteration process and the left new-constructed Hamiltonian has less eigen-components and is purer, which makes it easier to extract the remaining eigenstates. We need to analyze the properties related to the number of iterations. The Figure \ref{fig_K} shows the relationship between the number of iterations $k_{i}$ required to achieve chemical accuracy (0.0016 Hartree) and the energy-level difference. It can be found that a larger energy-level interval generally requires fewer iterations, but the highest energy state requires only one iteration ($k_{4}=1$), independent of the energy-level difference. We also plot the iterations times with bond length and find the maximum times $K=\textup{max}\{k_{i}\}$. We will show below that our FQESS algorithm can achieve faster convergence compared with other typical hybrid variational quantum algorithms for excited-states.

\begin{figure*}
\centering
\includegraphics[width=0.4\linewidth]{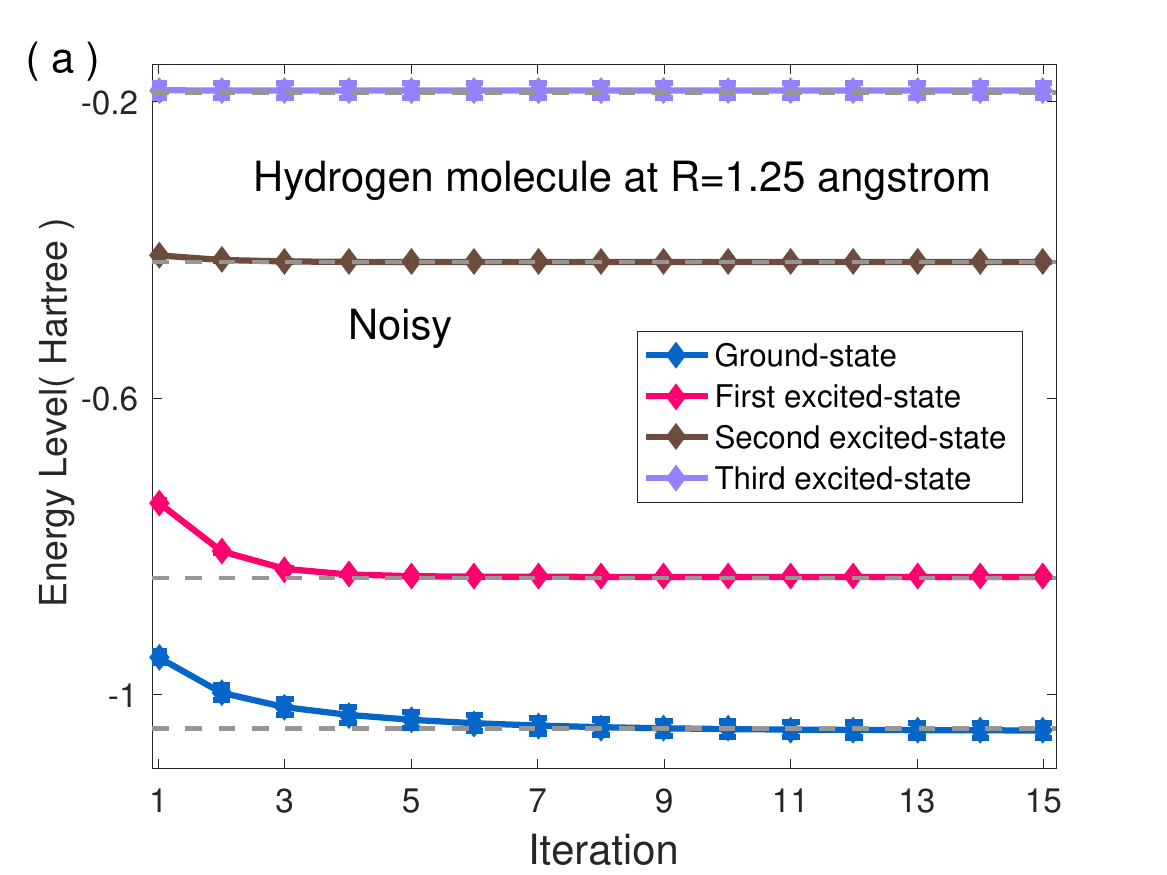}
\includegraphics[width=0.4\linewidth]{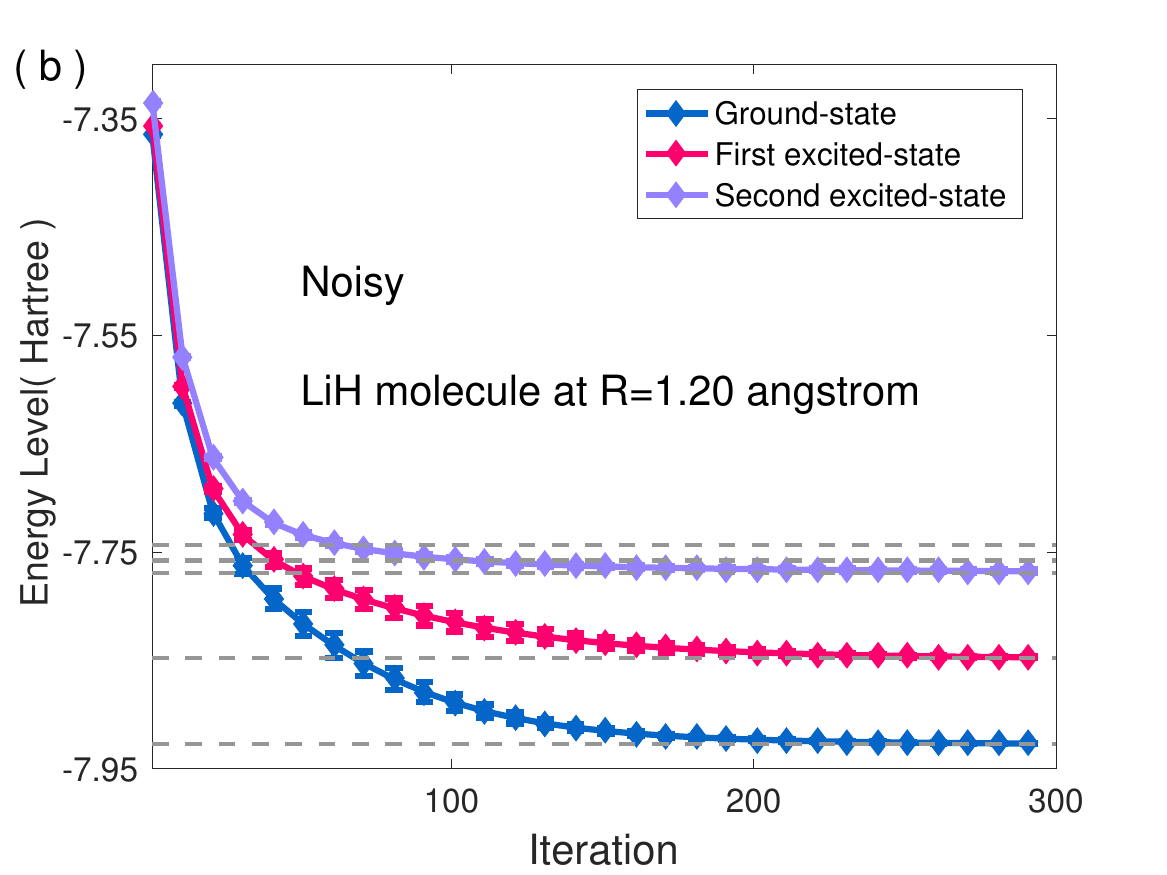}
\includegraphics[width=0.4\linewidth]{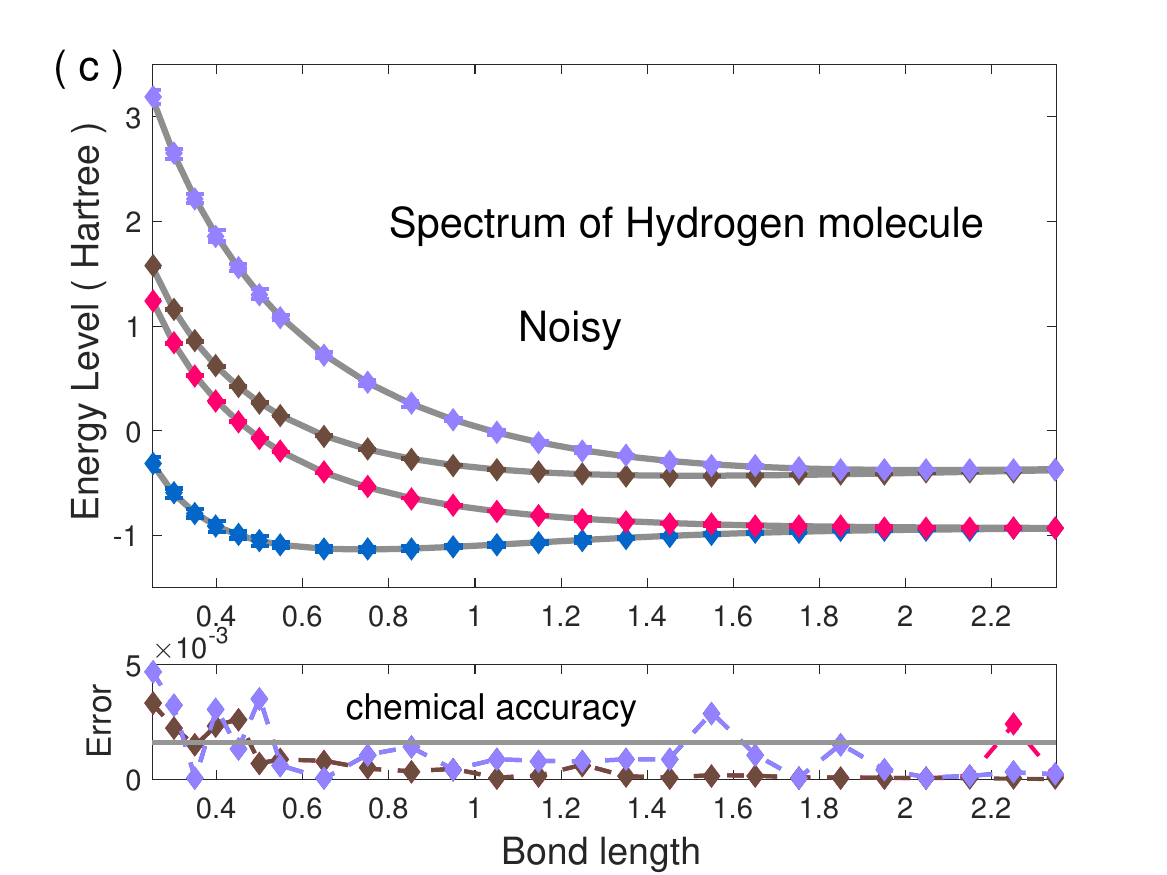}
\includegraphics[width=0.4\linewidth]{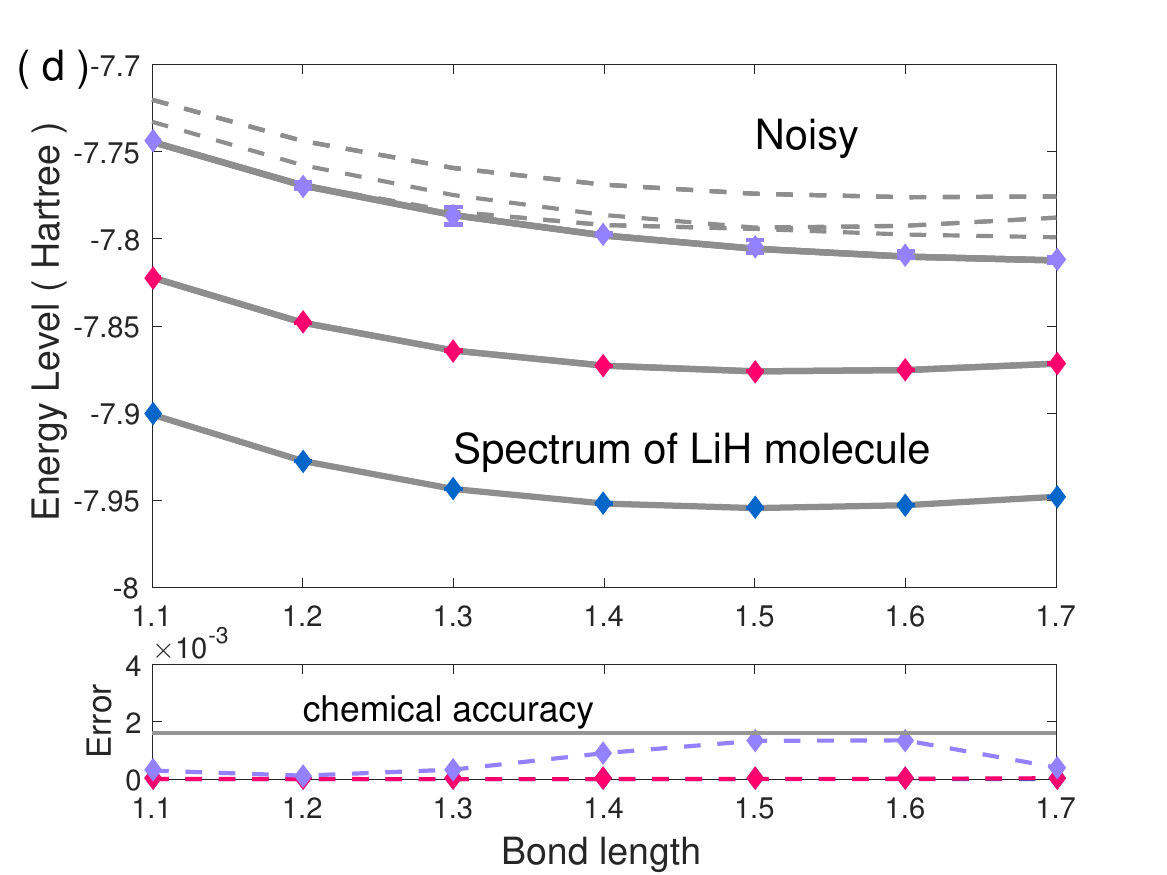}
\caption{A simulation of the hydrogen and LiH molecule with noise. (a,b) The iteration processes for different energy-levels at a specific bond length. (c,d) The whole spectrum for a range of inter-nuclear separations (unit in angstrom). The exact values are obtained by Hamiltonian diagonalization and plotted by lines. The numerical results are plotted with circles and the maximum deviation from the mean values are used as error bars. Errors between the numerical outputs and the theoretical expectations are shown in the bottom panel. }
\label{fig_iterationANDspectrum_noise} 
\end{figure*} 

We plot the whole spectrum of the hydrogen and LiH molecules for a wide range of inter-nuclear separations in Figure \ref{fig_spectrum_noiseless}. We set $k_{i}=600$ for better optimization results beyond chemical accuracy and the initial state for hydrogen is still $\ket{\psi_{0}}$ but for LiH molecule is $\ket{+}^{\otimes 6}$. By introducing suitable bias terms, we can find the energy-levels in turns, and the maximum differences for two molecules between the numerical outputs and the theoretical expectations are 0.000145 and 0.001203 Hartree, separately. Moreover, we study the effect of random noise term in the form $\sum_{i}^{n}\delta\alpha_{i}^{R}\sigma_{z}^{(i)}$, which is used to simulate decoherence noise with certain intensity \cite{wei2020full,wen2021variational}. In each optimization, we set iteration times as $k_{i}=600$ and these processes are repeated five times for average values, which is used as the output and the maximum deviation from the mean values are seen as error bars, as plotted in the Figure \ref{fig_iterationANDspectrum_noise}. We can conclude based on the output results that our algorithm is robust to the random noise and the fast convergence still can be achieved. The mean error of different eigenenergies between optimization results and theoretical expectations are 0.000978 and 0.000231 Hartree and the accuracy of most data points exceed the chemical accuracy. 

  \begin{figure*}
\centering
\includegraphics[width=0.74\linewidth]{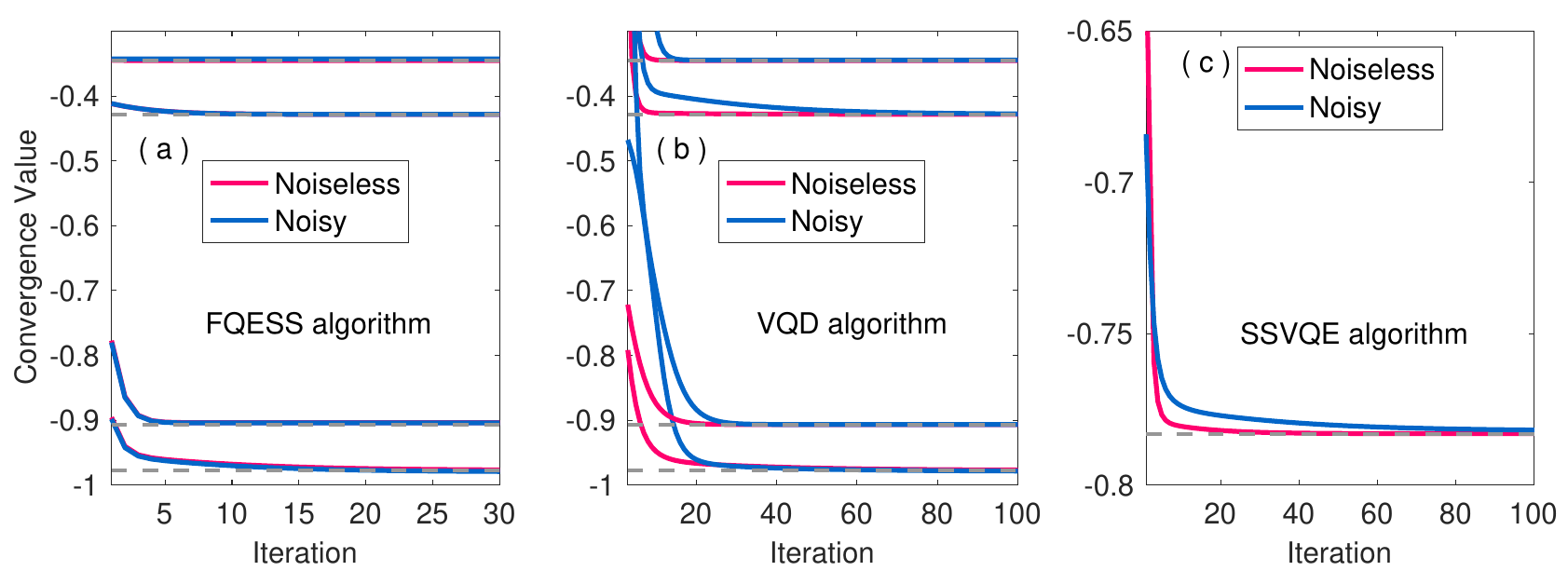}
\includegraphics[width=0.74\linewidth]{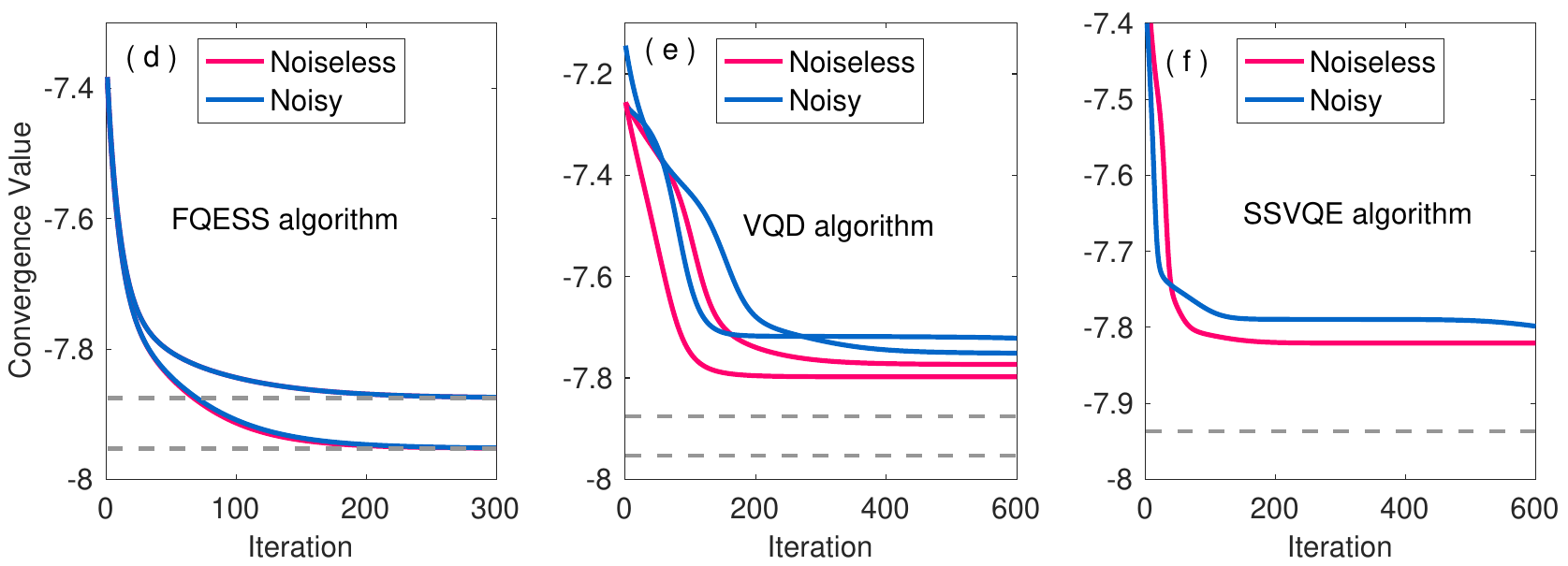}
\caption{Comparison of algorithms for excited-states between the FQESS, VQD and SSVQE algorithms with hydrogen (a-c) and LiH molecule (d-f). The performances of these algorithms under noiseless and 10\% noise are presented in red and blue lines, separately. The target convergence values labelled by gray dashed lines for FQESS and VQD algorithms are the eigenenergies of hydrogen molecule at $R=1.65$ angstrom and LiH molecule at $R=1.60$ angstrom, while target convergence value for SSVQE algorithm is a weighted average of energy eigenvalues.}
\label{fig_compare} 
\end{figure*} 

Furthermore, we compare the FQESS algorithm with another two typical hybrid variational algorithms for excited-states, i.e. VQD and SSVQE algorithm. The ansatz circuit for these two algorithms is the hardware-efficient ansatz \cite{kandala2017hardware} and the classical optimization algorithm used is Newton gradient descent algorithm. For hydrogen molecule, the initial state for VQD algorithm is same to that of FQESS algorithm when determining various eigenstates. The four initial orthogonal states in the SSVQE algorithm are chosen as the eigenstates of $P_{1}$ and the corresponding weights for them are $w_{1}=0.4$, $w_{2}=0.3$, $w_{3}=0.2$, $w_{4}=0.1$. So the target convergence value for SSVQE algorithm is $\sum_{i=1}^{4}w_{i}E_{i}$. For LiH molecule, the convergence values for FQESS and VQD algorithms are the eigenenergies of LiH molecule at $R=1.60$ angstrom, while target value for SSVQE algorithm is a weighted average ($w_{1}=0.8$, $w_{2}=0.2$) of the lowest two eigenvalues. The initial state for VQD and FQESS algorithm when determining various eigenstates is $\ket{+}^{\otimes 6}$ and the two initial orthogonal states for SSVQE algorithm are $\ket{+-}^{\otimes 3}$ and $\ket{-+}^{\otimes 3}$. The performances of these algorithms under noiseless and 10\% noise for hydrogen and LiH molecules are presented in the Figure \ref{fig_compare}. We can find that our FQESS algorithm shows stronger robustness against noise and the iteration times needed is less, which means it can achieve a faster convergence compared with the other two algorithms for excited-states. Although the two hybrid variational algorithms can be improved by changing the ansatz circuit or classical optimization process, these enhancements require pre-selection based on some prior knowledge, which is a potential challenge in practical applications, while the construction of FQESS algorithm only depends on the form of the target chemical Hamiltonian. The non-variational nature can also ensure that the FQESS algorithm converges to the target states along the direction of the fastest gradient descent, avoiding barren plateau. In addition, our algorithm can avoid the key challenges of overlap estimation realized by swap test in the VQD algorithm, and the limited number of solvable excited-states caused by the difficulty of simultaneous optimization of ansatz in the SSVQE algorithm.


\section{Experimental Demonstration} \label{Chapexp}

Here we demonstrate our FQESS algorithm with the real superconducting quantum computation chip on the Quafu quantum cloud platform \cite{ref:quafu}. Detail information about Quafu cloud platform can be found in the Appendix \ref{appendixQuafu}. As a proof-of-principle experimental demonstration, we consider the singlet and spatial symmetry of ground state in the hydrogen molecular, and then only two configurations are relevant in the calculation, i.e. the ground-state configuration and the double excitation configuration. Then the simplified Hamiltonian can be expressed as a two-dimensional matrix $H=\alpha_{0}+\alpha_{x}\sigma_{x}+\alpha_{z}\sigma_{z}$ in the Pauli basis with coefficients $\alpha_{0}=-1.04235$, $\alpha_{x}=0.1813$ and $\alpha_{z}=-0.78865$ \cite{du2010nmr}. Then a series of basic quantum logic gates can be applied as the quantum circuit introduced above and the encode operator on the ancillary qubit [$q0$] is $R_{y}(\beta)=[\cos(\beta/2),-\sin(\beta/2);\sin(\beta/2),\cos(\beta/2)]$ with $\beta=-2.6897$ for the ground state. The iteration process can be realized by introducing a rotation operator $R_{y}(\theta)$ on the work qubit [$q1$], which start from $\ket{+}$ state with $\theta=\pi/2$ initially.

\begin{figure}
\centering
\includegraphics[width=\linewidth]{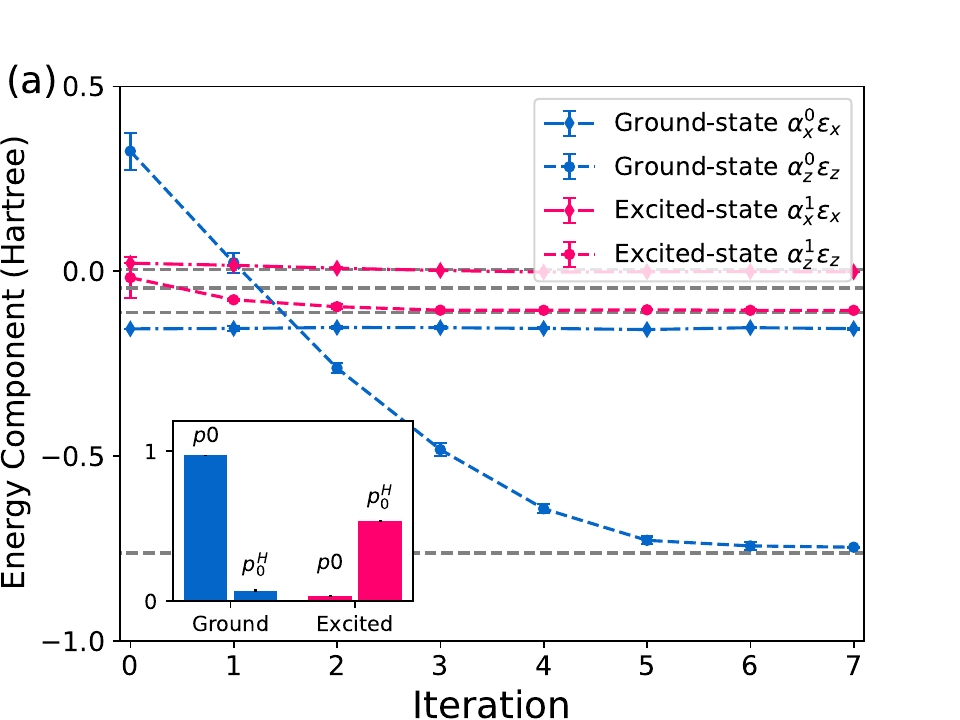}
\includegraphics[width=\linewidth]{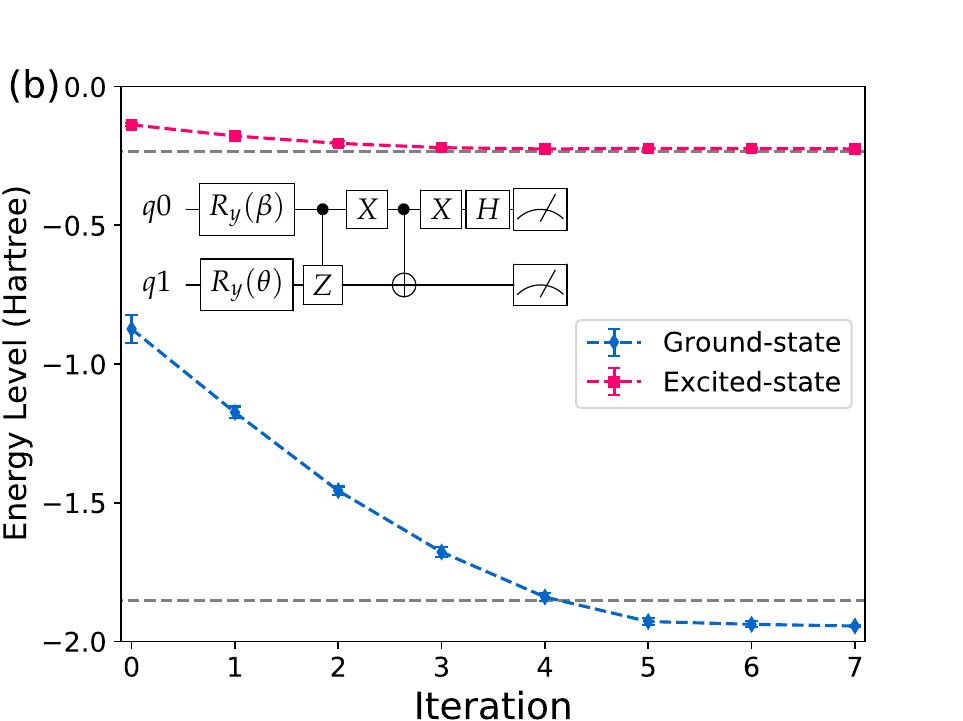}
\caption{(a) Energy components during iteration and final probability distribution for the ground- and excited-state. (b) Experimental quantum circuit and iteration results of FQESS algorithm with hydrogen molecular on the superconducting quantum computation chips.}
\label{fig_exp} 
\end{figure} 

The quantum processor will be called ten thousand times to obtain high-precision values in each setup and the experiment will be repeated three times to obtain error bars. As for the measurement, assume that the output quantum state of work system is $\ket{\phi_{\textup{I}}}$ and the experimentally-measurable distribution of $p0=\vert \inner{0}{\phi_{\textup{I}}} \vert^2$ and $p1=\vert \inner{1}{\phi_{\textup{I}}} \vert^2$  can be determined in the subspace of ancillary qubit. Then two energy components can be determined as $\epsilon_{0}=\inner{\phi_{\textup{I}}}{\phi_{\textup{I}}}=1$,  and $\epsilon_{z}=\bra{\phi_{\textup{I}}}\sigma_{z}\ket{\phi_{\textup{I}}}=p0-p1$. Because that $\epsilon_{x}=\bra{\phi_{\textup{I}}}\sigma_{x}\ket{\phi_{\textup{I}}}=\bra{\phi_{\textup{I}}}\mathcal{H}\sigma_{z}\mathcal{H}\ket{\phi_{\textup{I}}}$ where $\mathcal{H}$ represents Hardmard gate, we can repeat the same circuit but add $\mathcal{H}$ on work qubit to obtain $\ket{\phi_{\mathcal{H}}}=\mathcal{H}\ket{\phi_{\textup{I}}}$ with new probability distribution $p^\mathcal{H}_0$ and $p^\mathcal{H}_1$, and then $\epsilon_{x}=\bra{\phi_{\mathcal{H}}}\epsilon_{z}\ket{\phi_{\mathcal{H}}}=p^\mathcal{H}_0-p^\mathcal{H}_1$. Therefore, the energy can be reconstructed as  $E_{\textup{exp}}=\sum_{i=0,x,z}\alpha_{i}\epsilon_{i}$ experimentally. To realize the multi-step iteration, we need to repeat another same evolution process, but start with the output state in the last iteration. This can be realized by modifying the angle $\theta$ with formula $\theta=-2\arcsin(p1^{0.5})$. The experimental quantum circuit and iteration results of energy are plotted in the Figure \ref{fig_exp}, which show a good agreement with the theoretical expectations.


\section{Discussion} \label{Chapconclusion}

In summary, we have proposed a full quantum algorithm in circuit frame for the excited-states of quantum chemistry, termed FQESS algorithm. Compared with the hybrid variational algorithm, our method does not need the classical optimizer and all the calculations are performed on the quantum computer. The non-variational characteristic makes the algorithm converge to the target states along the direction of the fastest gradient descent, avoiding the barren plateau phenomenon. The output results of energy for the last energy-level can be used as the updating parameter for determining the next energy-level, and the only difference between different iterations for various eigenstates is the state preparation process of ancillary system, which can be simply realized by modifying the encoding operator. We present a numerical simulation with the hydrogen, LiH, H2O and NH3 molecules to demonstrate the feasibility and robustness of the algorithm. A proof-of-principle experiment has also been demonstrated on the practical superconducting quantum chip, and the results show a good agreement with theoretical expectations. Our algorithm fills the last step of solving quantum chemistry problems based on different algorithm frames and can be used as a generalized Hamiltonian diagonalization scheme on the future fault-tolerant quantum computers.


\section*{Acknowledgements} 

S. W. and G. L. are corresponding authors.
We acknowledge the support from the National Natural Science Foundation of China under Grants No. 12005015 and No. 11974205; the National Key R\&D Plan (2021YFB2801800); the National Key Research and Development Program of China (2017YFA0303700);  Beijing Advanced Innovation Center for Future Chip (ICFC). We gratefully acknowledge support from the Extreme Condition User Facility in Beijing and Quafu cloud platform for quantum computation. S. W. acknowledges the Beijing Nova Program under Grants No. 20230484345. H. F. acknowledge the support from the National Natural Science Foundation of China under Grants No. T2121001 and No. 92265207.  Z. W. acknowledge the support from China Postdoctoral Science Foundation No. 2022TQ0036, National Natural Science Foundation of China  No. 12247168 and No. 92265207. 





\appendix 

\section{Theory of the FQESS Algorithm}\label{FQESSbaisc}

The eigen-equation for an $n\times n$ matrix $A$ can be written as $Au^{(i)}=\lambda_{i}u^{(i)}$, where $\lambda_{i}$ is eigenvalue and $u^{(i)}$ is the corresponding eigenvector. For an arbitrarily selected vector $x^{(0)}$, it can always be decomposed in the eigenvector basis as $x^{(0)}=\sum_{i}^{n}a_{i}u^{(i)}$. Then we can multiply the matrix $A$ with $k$ times to $x^{(0)}$ and we have 
\begin{equation}
x^{(k)}=A^{k}x^{(0)}=\lambda_{1}^{k}\Big[ a_{1}u^{(1)}+\sum_{i=2}^{n}a_{i}(\frac{\lambda_{i}}{\lambda_{1}})^{k}u^{(i)}\Big]
\label{cm}
\end{equation}

\begin{figure*}
\centering
\includegraphics[width=0.49\linewidth]{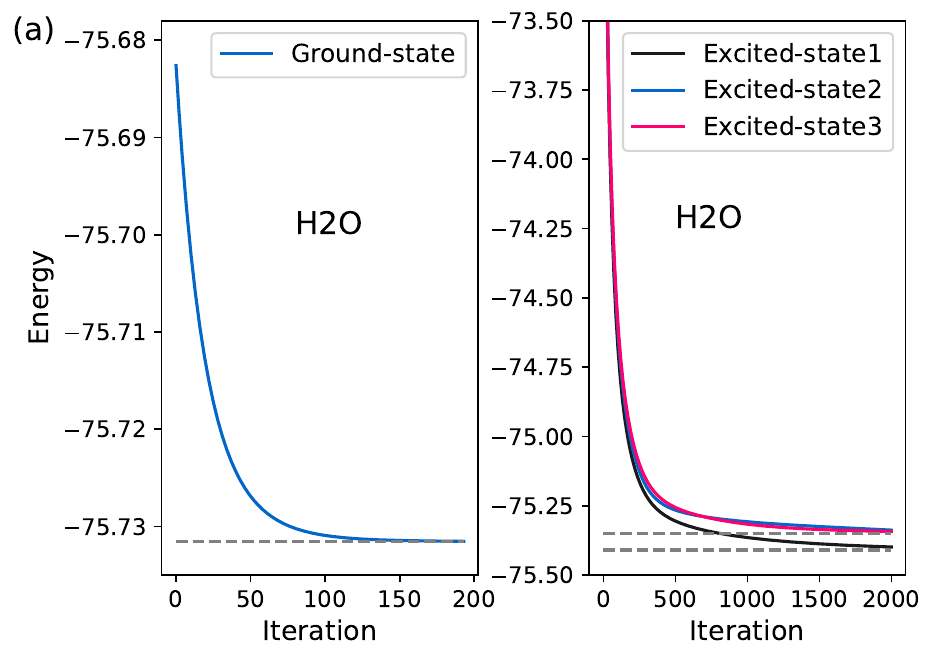}
\includegraphics[width=0.49\linewidth]{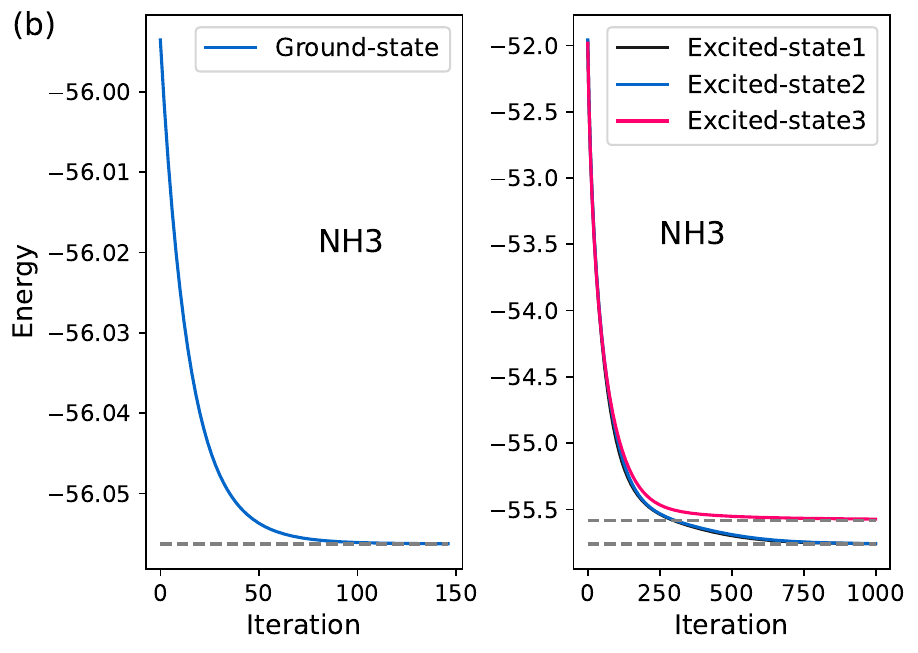}
\caption{Numerical simulations of the 12-qubit H2O (a) and 14-qubit NH3 (b) molecules. The left side in each panel shows the iteration processes of energy (unit in Hartree) for ground-states, while the right side shows that for excited-states. The target convergence energies obtained by Hamiltonian diagonalization are indicated by gray dashed lines.}
\label{fig_big} 
\end{figure*} 

Suppose that the absolute values of the eigenvalues satisfy the relation $\vert \lambda_{1} \vert > \vert \lambda_{2} \vert \ge \cdots \ge \vert \lambda_{n} \vert$, then $\lim_{k\to \infty} (\lambda_{i}/\lambda_{1})^{k}=0$. Therefore, the second item in equation (\ref{cm}) will vanish when we apply enough times of matrix $A$. Under the assumption that the initial vector has finite overlap with the targeted eigenvector ($a_{1}\ne0$), the remaining term will be proportional to the eigenvector $u^{(1)}$ with the biggest eigenvalue, except for a coefficient, which can be eliminated by normalization. The process presented above is termed power iteration method, which is a common iterative method to calculate the maximum absolute value of eigenvalue and the corresponding eigenvector (principal component) of the matrix \cite{panju2011iterative}. This intuitive and elegant method works well for the eigenvector estimation of the large and sparse matrices. We take this method as the basic eigen-component extracter of the chemical Hamiltonian in the FQESS algorithm.

In addition, the method involved in the FQESS algorithm can also be interpreted from the perspective of quantum gradient descent \cite{wei2020full}.  The target function can be expressed as a quadratic optimization problem as $f(\ket{X})=\bra{X}H\ket{X}$, and then the state evolving along the direction of the gradient of target function can be expressed as 
\begin{equation}
\ket{X^{t+1}}=\ket{X^{t}}-\gamma\nabla f(\ket{X^{t}})=(I-2\gamma H)\ket{X^{t}}
\label{gradient}
\end{equation}
where $\gamma$ represents the learning rate. If the bias parameter is set as $\lambda_{0}=1/2\gamma$ and ignore an unimportant multiplying factor, the gradient operator $U_{g}=(I-2\gamma H)$ will be same to the operator $U_{i}$ in equation (\ref{Ui}), both of which are linear combinations of unitary operators. Therefore, the iteration process of FQESS algorithm can also be regarded as the process of quantum state converging to the specific eigenstate along the gradient direction of objective function.


\section{Simulation of H2O and NH3 molecules}\label{appendixbigger}

To demonstrate the performance of the FQESS algorithm in larger systems, we offer the numerical simulations with 12-qubit H2O and 14-qubit NH3 molecules in STO-6G basis set here, providing additional support for the effectiveness of our algorithm. For the ground-states, we start from Hartree-Fock states, and use random initial quantum states in excited-states for better convergence. Other setup is same to the simulation in the hydrogen and LiH molecules. As shown in the Figure \ref{fig_big}, the FQESS algorithm can well converge to the target solution of ground-states and excited-states, indicating that the algorithm is feasible to solve the energy spectrum in larger molecular systems. Compared with the exact solutions obtained by Hamiltonian diagonalization, the error of ground-state energy for H2O (NH3) molecule is 0.000043 (0.000029) Hartree,  and the average error of excited-states is 0.001163 (0.000399) Hartree, both reaching the chemical accuracy.


\section{Quafu Quantum Cloud Platform}\label{appendixQuafu}

\begin{figure}
\centering
\includegraphics[width=0.35\textwidth]{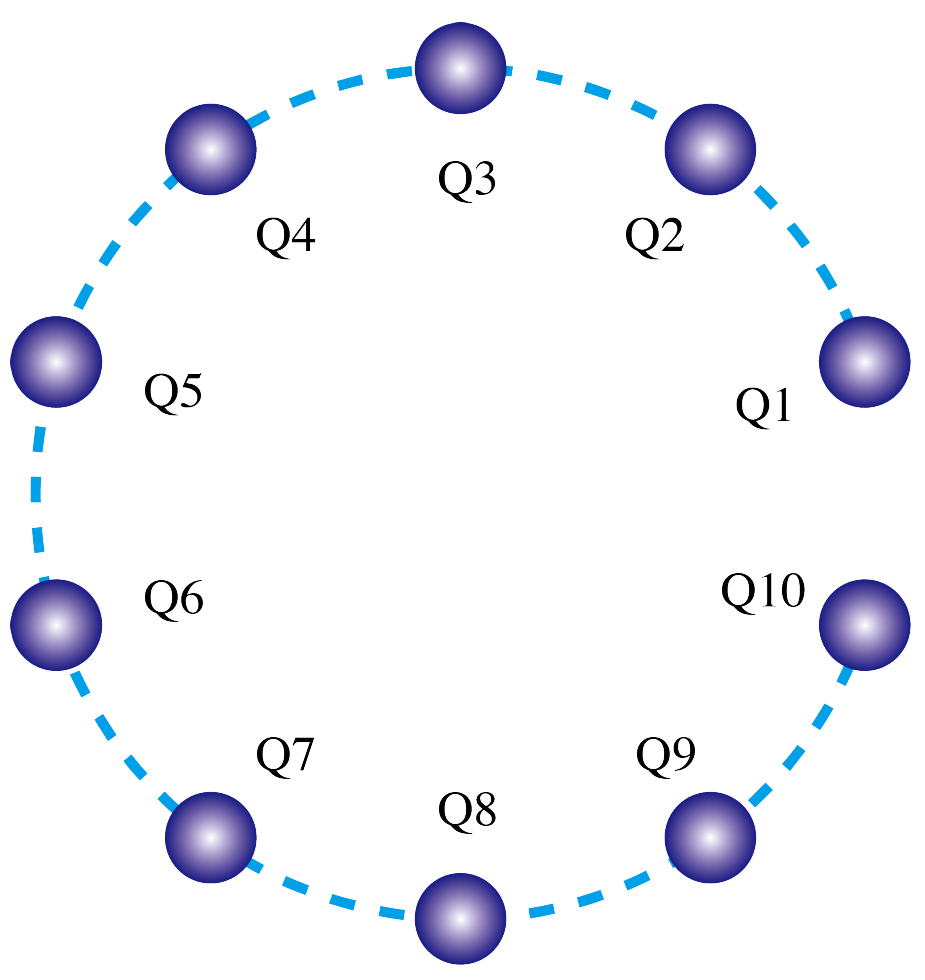}
\caption{The topological structure of quantum processor P-10. Each qubit capacitively coupled to its nearest-neighbors.}
\label{fig_p10}
\end{figure}

Quafu is an open cloud platform for quantum computation \cite{ref:quafu}, which provides four specifications of superconducting quantum processors currently. Three of them support  general quantum logical gates, which are 10-qubits and 18-qubits processors with one-dimensional chain structure named P-10 and P-18, and a 50$+$ qubits processor with a two-dimensional honeycomb structure named P-50. In this work, we use the first two qubits of P-10 quantum processor, whose topological structure is shown in the Figure~\ref{fig_p10}. The processor consists of ten transmon qubits ($Q_1-Q_{10}$) arrayed in a row, with each qubit capacitively coupled to its nearest-neighbors. Each transmon qubit can be modulated in frequency from about 4 to 5.7 GHz and excited to the excited state individually. All qubits can be probed though a common transmission line connected to their own readout resonators. The qubit parameters and coherence performance can be found in the Table \ref{tab:1}. The idle frequencies of each qubit ($\omega^{10}_j$) are designed to reduce residual coupling strength from other qubits.

\begin{table*}
    \centering
    \setlength{\tabcolsep}{2.5mm}{
    \begin{tabular}{ccccccccccc} 
        \toprule
         qubit&$Q_{1}$&$Q_{2}$&$Q_{3}$&$Q_{4}$&$Q_{5}$&$Q_{6}$&$Q_{7}$&$Q_{8}$&$Q_{9}$&$Q_{10}$\\
        \midrule
         $\omega^{s}_{j}/2\pi$ (GHz)  &5.536 &5.069 &5.660 &4.742 &5.528 &4.929 &5.451 &4.920 &5.540 &4.960 \\
        $\omega^{10}_{j}/2\pi$  (GHz)    &5.456 &4.424 &5.606 &4.327 &5.473 &4.412 &5.392 &4.319 &5.490 &4.442 \\
        $\omega^{r}_{j}/2\pi$  (GHz) & 5.088 &4.702 &5.606 &4.466 &5.300 &4.804 &5.177 &4.697 &5.474 &4.819 \\
        $\eta_{j}/2\pi$   (GHz) & 0.250 &0.207 &0.251 &0.206 &0.251 &0.203 &0.252 &0.204 &0.246 &0.208 \\
        $g_{j,j+1}/2\pi$ (MHz) & 12.07 &11.58 &10.92 &10.84 &11.56 &10.00 &11.74 &11.70 &11.69 & - \\
        $T_{1,j}$ (us)    &20.0 &52.5 &15.9 &16.3 &36.9 &44.4 &30.8 &77.7 &22.8 &25.0 \\
        $T_{2,j}^{*}$ (us) &8.60 &1.48 &9.11 &2.10 &12.8 &2.73 &15.7 &1.88 &4.49 &2.05 \\
        $F_{0,j}$ (\%) & 98.90 &98.32 &98.67 &95.30 &97.00 &95.47 &97.00 &96.37 &98.33 &97.13 \\
        $F_{1,j}$ (\%) & 92.90 &92.30 &92.97 &91.53 &86.17 &87.93 &93.40 &93.37 &94.63 &92.07 \\
        $F_{j,j+1}$ (\%) &94.2 &97.8 &96.6 &97.3 &96.8 &97.0 &94.5 &93.2 &96.0 & -\\
        \bottomrule
    \end{tabular}}
    \caption{Device parameters. $\omega^{s}_{j}$ shows the maximum frequency of $Q_{j}$. $\omega^{10}_{j}$ corresponds to the idle frequency of $Q_{j}$. $\omega^{r}_{j}$ shows the resonant frequency of $Q_{j}$ during readout. $\eta_j$ corresponds to  the anharmonicity of $Q_{j}$. $g_{j,j+1}$ is the coupling strength between nearest-neighbor qubits. $T_{1,j}$ and $T^{*}_{2,j}$ represent the relaxation time and coherence time of $Q_{j}$. $F_{0,j}$ and $F_{1,j}$ are readout fidelities of $Q_{j}$ in $|0\rangle$ and $|1\rangle$. $F_{j,j+1}$ represents the fidelity of CZ gate composed of $Q_{i}$ and $Q_{j}$, which is obtianed by randomized benchmarking.}
\label{tab:1}
\end{table*}


\bibliographystyle{unsrtnat}
\bibliography{FQESS_wen.bib}

\begin{thebibliography}{49}
\providecommand{\natexlab}[1]{#1}
\providecommand{\url}[1]{\texttt{#1}}
\expandafter\ifx\csname urlstyle\endcsname\relax
  \providecommand{\doi}[1]{doi: #1}\else
  \providecommand{\doi}{doi: \begingroup \urlstyle{rm}\Url}\fi

\bibitem[Benioff(1980)]{benioff1980computer}
Paul Benioff.
\newblock The computer as a physical system: A microscopic quantum mechanical
  hamiltonian model of computers as represented by turing machines.
\newblock \emph{Journal of statistical physics}, 22\penalty0 (5):\penalty0
  563--591, 1980.
\newblock \doi{10.1007/BF01011339}.

\bibitem[Feynman(1982)]{feynman2018simulating}
Richard~P Feynman.
\newblock Simulating physics with computers.
\newblock \emph{Int J Theor Phys}, 21\penalty0 (1):\penalty0 467--488, 1982.
\newblock \doi{10.1007/BF02650179}.

\bibitem[Shor(1999)]{shor1999polynomial}
Peter~W Shor.
\newblock Polynomial-time algorithms for prime factorization and discrete
  logarithms on a quantum computer.
\newblock \emph{SIAM review}, 41\penalty0 (2):\penalty0 303--332, 1999.
\newblock \doi{10.1137/S0036144598347011}.

\bibitem[Grover(1997)]{grover1997quantum}
Lov~K Grover.
\newblock Quantum mechanics helps in searching for a needle in a haystack.
\newblock \emph{Physical review letters}, 79\penalty0 (2):\penalty0 325, 1997.
\newblock \doi{10.1103/PhysRevLett.79.325}.

\bibitem[Long et~al.(1999)Long, Li, Zhang, and Niu]{long1999phase}
Gui~Lu Long, Yan~Song Li, Wei~Lin Zhang, and Li~Niu.
\newblock Phase matching in quantum searching.
\newblock \emph{Physics Letters A}, 262\penalty0 (1):\penalty0 27--34, 1999.
\newblock \doi{10.1016/S0375-9601(99)00631-3}.

\bibitem[Harrow et~al.(2009)Harrow, Hassidim, and Lloyd]{harrow2009quantum}
Aram~W Harrow, Avinatan Hassidim, and Seth Lloyd.
\newblock Quantum algorithm for linear systems of equations.
\newblock \emph{Physical review letters}, 103\penalty0 (15):\penalty0 150502,
  2009.
\newblock \doi{10.1103/PhysRevLett.103.150502}.

\bibitem[Suba{\c{s}}{\i} et~al.(2019)Suba{\c{s}}{\i}, Somma, and
  Orsucci]{subacsi2019quantum}
Yi{\u{g}}it Suba{\c{s}}{\i}, Rolando~D Somma, and Davide Orsucci.
\newblock Quantum algorithms for systems of linear equations inspired by
  adiabatic quantum computing.
\newblock \emph{Physical review letters}, 122\penalty0 (6):\penalty0 060504,
  2019.
\newblock \doi{10.1103/PhysRevLett.122.060504}.

\bibitem[Cao et~al.(2019)Cao, Romero, Olson, Degroote, Johnson, Kieferov{\'a},
  Kivlichan, Menke, Peropadre, Sawaya, et~al.]{cao2019quantum}
Yudong Cao, Jonathan Romero, Jonathan~P Olson, Matthias Degroote, Peter~D
  Johnson, M{\'a}ria Kieferov{\'a}, Ian~D Kivlichan, Tim Menke, Borja
  Peropadre, Nicolas~PD Sawaya, et~al.
\newblock Quantum chemistry in the age of quantum computing.
\newblock \emph{Chemical reviews}, 119\penalty0 (19):\penalty0 10856--10915,
  2019.
\newblock \doi{10.1021/acs.chemrev.8b00803}.

\bibitem[McArdle et~al.(2020)McArdle, Endo, Aspuru-Guzik, Benjamin, and
  Yuan]{mcardle2020quantum}
Sam McArdle, Suguru Endo, Al{\'a}n Aspuru-Guzik, Simon~C Benjamin, and Xiao
  Yuan.
\newblock Quantum computational chemistry.
\newblock \emph{Reviews of Modern Physics}, 92\penalty0 (1):\penalty0 015003,
  2020.
\newblock \doi{10.1103/RevModPhys.92.015003}.

\bibitem[Bauer et~al.(2020)Bauer, Bravyi, Motta, and Chan]{bauer2020quantum}
Bela Bauer, Sergey Bravyi, Mario Motta, and Garnet Kin-Lic Chan.
\newblock Quantum algorithms for quantum chemistry and quantum materials
  science.
\newblock \emph{Chemical Reviews}, 120\penalty0 (22):\penalty0 12685--12717,
  2020.
\newblock \doi{10.1021/acs.chemrev.9b00829}.

\bibitem[Peruzzo et~al.(2014)Peruzzo, McClean, Shadbolt, Yung, Zhou, Love,
  Aspuru-Guzik, and O’brien]{peruzzo2014variational}
Alberto Peruzzo, Jarrod McClean, Peter Shadbolt, Man-Hong Yung, Xiao-Qi Zhou,
  Peter~J Love, Al{\'a}n Aspuru-Guzik, and Jeremy~L O’brien.
\newblock A variational eigenvalue solver on a photonic quantum processor.
\newblock \emph{Nature communications}, 5\penalty0 (1):\penalty0 1--7, 2014.
\newblock \doi{10.1038/ncomms5213}.

\bibitem[O’Malley et~al.(2016)O’Malley, Babbush, Kivlichan, Romero,
  McClean, Barends, Kelly, Roushan, Tranter, Ding, et~al.]{o2016scalable}
Peter~JJ O’Malley, Ryan Babbush, Ian~D Kivlichan, Jonathan Romero, Jarrod~R
  McClean, Rami Barends, Julian Kelly, Pedram Roushan, Andrew Tranter, Nan
  Ding, et~al.
\newblock Scalable quantum simulation of molecular energies.
\newblock \emph{Physical Review X}, 6\penalty0 (3):\penalty0 031007, 2016.
\newblock \doi{10.1103/PhysRevX.6.031007}.

\bibitem[Kandala et~al.(2017)Kandala, Mezzacapo, Temme, Takita, Brink, Chow,
  and Gambetta]{kandala2017hardware}
Abhinav Kandala, Antonio Mezzacapo, Kristan Temme, Maika Takita, Markus Brink,
  Jerry~M Chow, and Jay~M Gambetta.
\newblock Hardware-efficient variational quantum eigensolver for small
  molecules and quantum magnets.
\newblock \emph{Nature}, 549\penalty0 (7671):\penalty0 242--246, 2017.
\newblock \doi{10.1038/nature23879}.

\bibitem[Cerezo et~al.(2021)Cerezo, Arrasmith, Babbush, Benjamin, Endo, Fujii,
  McClean, Mitarai, Yuan, Cincio, et~al.]{cerezo2021variational}
Marco Cerezo, Andrew Arrasmith, Ryan Babbush, Simon~C Benjamin, Suguru Endo,
  Keisuke Fujii, Jarrod~R McClean, Kosuke Mitarai, Xiao Yuan, Lukasz Cincio,
  et~al.
\newblock Variational quantum algorithms.
\newblock \emph{Nature Reviews Physics}, pages 1--20, 2021.
\newblock \doi{10.1038/s42254-021-00348-9}.

\bibitem[Bonet-Monroig et~al.(2018)Bonet-Monroig, Sagastizabal, Singh, and
  O'Brien]{bonet2018low}
Xavi Bonet-Monroig, Ramiro Sagastizabal, M~Singh, and TE~O'Brien.
\newblock Low-cost error mitigation by symmetry verification.
\newblock \emph{Physical Review A}, 98\penalty0 (6):\penalty0 062339, 2018.
\newblock \doi{10.1103/PhysRevA.98.062339}.

\bibitem[Grimsley et~al.(2019)Grimsley, Economou, Barnes, and
  Mayhall]{grimsley2019adaptive}
Harper~R Grimsley, Sophia~E Economou, Edwin Barnes, and Nicholas~J Mayhall.
\newblock An adaptive variational algorithm for exact molecular simulations on
  a quantum computer.
\newblock \emph{Nature communications}, 10\penalty0 (1):\penalty0 1--9, 2019.
\newblock \doi{10.1038/s41467-019-10988-2}.

\bibitem[Tang et~al.(2021)Tang, Shkolnikov, Barron, Grimsley, Mayhall, Barnes,
  and Economou]{tang2021qubit}
Ho~Lun Tang, VO~Shkolnikov, George~S Barron, Harper~R Grimsley, Nicholas~J
  Mayhall, Edwin Barnes, and Sophia~E Economou.
\newblock qubit-adapt-vqe: An adaptive algorithm for constructing
  hardware-efficient ans{\"a}tze on a quantum processor.
\newblock \emph{PRX Quantum}, 2\penalty0 (2):\penalty0 020310, 2021.
\newblock \doi{10.1103/PRXQuantum.2.020310}.

\bibitem[Ostaszewski et~al.(2021)Ostaszewski, Grant, and
  Benedetti]{ostaszewski2021structure}
Mateusz Ostaszewski, Edward Grant, and Marcello Benedetti.
\newblock Structure optimization for parameterized quantum circuits.
\newblock \emph{Quantum}, 5:\penalty0 391, 2021.
\newblock \doi{10.22331/q-2021-01-28-391}.

\bibitem[Wei et~al.(2020)Wei, Li, and Long]{wei2020full}
Shijie Wei, Hang Li, and GuiLu Long.
\newblock A full quantum eigensolver for quantum chemistry simulations.
\newblock \emph{Research}, 2020, 2020.
\newblock \doi{10.34133/2020/1486935}.

\bibitem[Rebentrost et~al.(2019)Rebentrost, Schuld, Wossnig, Petruccione, and
  Lloyd]{rebentrost2019quantum}
Patrick Rebentrost, Maria Schuld, Leonard Wossnig, Francesco Petruccione, and
  Seth Lloyd.
\newblock Quantum gradient descent and newton’s method for constrained
  polynomial optimization.
\newblock \emph{New Journal of Physics}, 21\penalty0 (7):\penalty0 073023,
  2019.
\newblock \doi{10.1088/1367-2630/ab2a9e}.

\bibitem[Higgott et~al.(2019)Higgott, Wang, and
  Brierley]{higgott2019variational}
Oscar Higgott, Daochen Wang, and Stephen Brierley.
\newblock Variational quantum computation of excited states.
\newblock \emph{Quantum}, 3:\penalty0 156, 2019.
\newblock \doi{10.22331/q-2019-07-01-156}.

\bibitem[Jones et~al.(2019)Jones, Endo, McArdle, Yuan, and
  Benjamin]{jones2019variational}
Tyson Jones, Suguru Endo, Sam McArdle, Xiao Yuan, and Simon~C Benjamin.
\newblock Variational quantum algorithms for discovering hamiltonian spectra.
\newblock \emph{Physical Review A}, 99\penalty0 (6):\penalty0 062304, 2019.
\newblock \doi{10.1103/PhysRevA.99.062304}.

\bibitem[Nakanishi et~al.(2019)Nakanishi, Mitarai, and
  Fujii]{nakanishi2019subspace}
Ken~M Nakanishi, Kosuke Mitarai, and Keisuke Fujii.
\newblock Subspace-search variational quantum eigensolver for excited states.
\newblock \emph{Physical Review Research}, 1\penalty0 (3):\penalty0 033062,
  2019.
\newblock \doi{10.1103/PhysRevResearch.1.033062}.

\bibitem[Parrish et~al.(2019)Parrish, Hohenstein, McMahon, and
  Mart{\'\i}nez]{parrish2019quantum}
Robert~M Parrish, Edward~G Hohenstein, Peter~L McMahon, and Todd~J
  Mart{\'\i}nez.
\newblock Quantum computation of electronic transitions using a variational
  quantum eigensolver.
\newblock \emph{Physical review letters}, 122\penalty0 (23):\penalty0 230401,
  2019.
\newblock \doi{10.1103/PhysRevLett.122.230401}.

\bibitem[McClean et~al.(2017)McClean, Kimchi-Schwartz, Carter, and
  De~Jong]{mcclean2017hybrid}
Jarrod~R McClean, Mollie~E Kimchi-Schwartz, Jonathan Carter, and Wibe~A
  De~Jong.
\newblock Hybrid quantum-classical hierarchy for mitigation of decoherence and
  determination of excited states.
\newblock \emph{Physical Review A}, 95\penalty0 (4):\penalty0 042308, 2017.
\newblock \doi{10.1103/PhysRevA.95.042308}.

\bibitem[Colless et~al.(2018)Colless, Ramasesh, Dahlen, Blok, Kimchi-Schwartz,
  McClean, Carter, de~Jong, and Siddiqi]{colless2018computation}
James~I Colless, Vinay~V Ramasesh, Dar Dahlen, Machiel~S Blok, Mollie~E
  Kimchi-Schwartz, Jarrod~R McClean, Jonathan Carter, Wibe~A de~Jong, and Irfan
  Siddiqi.
\newblock Computation of molecular spectra on a quantum processor with an
  error-resilient algorithm.
\newblock \emph{Physical Review X}, 8\penalty0 (1):\penalty0 011021, 2018.
\newblock \doi{10.1103/PhysRevX.8.011021}.

\bibitem[Jouzdani et~al.(2019)Jouzdani, Bringuier, and
  Kostuk]{jouzdani2019method}
Pejman Jouzdani, Stefan Bringuier, and Mark Kostuk.
\newblock A method of determining excited-states for quantum computation.
\newblock \emph{arXiv preprint arXiv:1908.05238}, 2019.
\newblock \doi{10.48550/arXiv.1908.05238}.

\bibitem[Ollitrault et~al.(2020)Ollitrault, Kandala, Chen, Barkoutsos,
  Mezzacapo, Pistoia, Sheldon, Woerner, Gambetta, and
  Tavernelli]{ollitrault2020quantum}
Pauline~J Ollitrault, Abhinav Kandala, Chun-Fu Chen, Panagiotis~Kl Barkoutsos,
  Antonio Mezzacapo, Marco Pistoia, Sarah Sheldon, Stefan Woerner, Jay~M
  Gambetta, and Ivano Tavernelli.
\newblock Quantum equation of motion for computing molecular excitation
  energies on a noisy quantum processor.
\newblock \emph{Physical Review Research}, 2\penalty0 (4):\penalty0 043140,
  2020.
\newblock \doi{10.1103/PhysRevResearch.2.043140}.

\bibitem[Zhang et~al.(2022)Zhang, Chen, Yuan, and Yin]{zhang2020variational}
Dan-Bo Zhang, Bin-Lin Chen, Zhan-Hao Yuan, and Tao Yin.
\newblock Variational quantum eigensolvers by variance minimization.
\newblock \emph{Chinese Physics B}, 31\penalty0 (12):\penalty0 120301, 2022.
\newblock \doi{10.1088/1674-1056/ac8a8d}.

\bibitem[Yalouz et~al.(2022)Yalouz, Koridon, Senjean, Lasorne, Buda, and
  Visscher]{yalouz2021analytical}
Saad Yalouz, Emiel Koridon, Bruno Senjean, Benjamin Lasorne, Francesco Buda,
  and Lucas Visscher.
\newblock Analytical nonadiabatic couplings and gradients within the
  state-averaged orbital-optimized variational quantum eigensolver.
\newblock \emph{Journal of chemical theory and computation}, 18\penalty0
  (2):\penalty0 776--794, 2022.
\newblock \doi{10.1021/acs.jctc.1c00995}.

\bibitem[Wen et~al.(2021)Wen, Lv, Yung, and Long]{wen2021variational}
Jingwei Wen, Dingshun Lv, Man-Hong Yung, and Gui-Lu Long.
\newblock Variational quantum packaged deflation for arbitrary excited states.
\newblock \emph{Quantum Engineering}, page e80, 2021.
\newblock \doi{10.1002/que2.80}.

\bibitem[Jordan and Wigner(1993)]{jordan1993paulische}
Pascual Jordan and Eugene~Paul Wigner.
\newblock {\"u}ber das paulische {\"a}quivalenzverbot.
\newblock In \emph{The Collected Works of Eugene Paul Wigner}, pages 109--129.
  Springer, 1993.
\newblock \doi{10.1007/978-3-662-02781-3_9}.

\bibitem[Bravyi and Kitaev(2002)]{bravyi2002fermionic}
Sergey~B Bravyi and Alexei~Yu Kitaev.
\newblock Fermionic quantum computation.
\newblock \emph{Annals of Physics}, 298\penalty0 (1):\penalty0 210--226, 2002.
\newblock \doi{10.1006/aphy.2002.6254}.

\bibitem[Gui-Lu(2006)]{gui2006general}
Long Gui-Lu.
\newblock General quantum interference principle and duality computer.
\newblock \emph{Communications in Theoretical Physics}, 45\penalty0
  (5):\penalty0 825, 2006.
\newblock \doi{10.1088/0253-6102/45/5/013}.

\bibitem[Gui-Lu and Yang(2008)]{gui2008duality}
Long Gui-Lu and Liu Yang.
\newblock Duality computing in quantum computers.
\newblock \emph{Communications in Theoretical Physics}, 50\penalty0
  (6):\penalty0 1303, 2008.
\newblock \doi{10.1088/0253-6102/50/6/11}.

\bibitem[Gui-Lu et~al.(2009)Gui-Lu, Yang, and Chuan]{gui2009allowable}
Long Gui-Lu, Liu Yang, and Wang Chuan.
\newblock Allowable generalized quantum gates.
\newblock \emph{Communications in Theoretical Physics}, 51\penalty0
  (1):\penalty0 65, 2009.
\newblock \doi{10.1088/0253-6102/51/1/13}.

\bibitem[Childs and Wiebe(2012)]{childs2012hamiltonian}
Andrew~M Childs and Nathan Wiebe.
\newblock Hamiltonian simulation using linear combinations of unitary
  operations.
\newblock \emph{arXiv preprint arXiv:1202.5822}, 2012.
\newblock \doi{10.48550/arXiv.1202.5822}.

\bibitem[Wen et~al.(2019)Wen, Zheng, Kong, Wei, Xin, and
  Long]{wen2019experimental}
Jingwei Wen, Chao Zheng, Xiangyu Kong, Shijie Wei, Tao Xin, and Guilu Long.
\newblock Experimental demonstration of a digital quantum simulation of a
  general $\mathcal{PT}$-symmetric system.
\newblock \emph{Physical Review A}, 99\penalty0 (6):\penalty0 062122, 2019.
\newblock \doi{10.1103/PhysRevA.99.062122}.

\bibitem[Wen et~al.(2020)Wen, Qin, Zheng, Wei, Kong, Xin, and
  Long]{wen2020observation}
Jingwei Wen, Guoqing Qin, Chao Zheng, Shijie Wei, Xiangyu Kong, Tao Xin, and
  Guilu Long.
\newblock Observation of information flow in the anti-$\mathcal{PT}$-symmetric
  system with nuclear spins.
\newblock \emph{npj Quantum Information}, 6\penalty0 (1):\penalty0 1--7, 2020.
\newblock \doi{10.1038/s41534-020-0258-4}.

\bibitem[Long and Sun(2001)]{long2001efficient}
Gui-Lu Long and Yang Sun.
\newblock Efficient scheme for initializing a quantum register with an
  arbitrary superposed state.
\newblock \emph{Physical Review A}, 64\penalty0 (1):\penalty0 014303, 2001.
\newblock \doi{10.1103/PhysRevA.64.014303}.

\bibitem[Giovannetti et~al.(2008)Giovannetti, Lloyd, and
  Maccone]{giovannetti2008quantum}
Vittorio Giovannetti, Seth Lloyd, and Lorenzo Maccone.
\newblock Quantum random access memory.
\newblock \emph{Physical review letters}, 100\penalty0 (16):\penalty0 160501,
  2008.
\newblock \doi{10.1103/PhysRevLett.100.160501}.

\bibitem[Brassard et~al.(2002)Brassard, Hoyer, Mosca, and
  Tapp]{brassard2002quantum}
Gilles Brassard, Peter Hoyer, Michele Mosca, and Alain Tapp.
\newblock Quantum amplitude amplification and estimation.
\newblock \emph{Contemporary Mathematics}, 305:\penalty0 53--74, 2002.
\newblock \doi{10.1090/conm/305/05215}.

\bibitem[Berry et~al.(2015)Berry, Childs, Cleve, Kothari, and
  Somma]{berry2015simulating}
Dominic~W Berry, Andrew~M Childs, Richard Cleve, Robin Kothari, and Rolando~D
  Somma.
\newblock Simulating hamiltonian dynamics with a truncated taylor series.
\newblock \emph{Physical review letters}, 114\penalty0 (9):\penalty0 090502,
  2015.
\newblock \doi{10.1103/PhysRevLett.114.090502}.

\bibitem[Xin et~al.(2017)Xin, Wei, Pedernales, Solano, and
  Long]{xin2017quantum}
Tao Xin, Shi-Jie Wei, Julen~S Pedernales, Enrique Solano, and Gui-Lu Long.
\newblock Quantum simulation of quantum channels in nuclear magnetic resonance.
\newblock \emph{Physical Review A}, 96\penalty0 (6):\penalty0 062303, 2017.
\newblock \doi{10.1103/PhysRevA.96.062303}.

\bibitem[Wei et~al.(2018)Wei, Xin, and Long]{wei2018efficient}
Shi-Jie Wei, Tao Xin, and Gui-Lu Long.
\newblock Efficient universal quantum channel simulation in ibm’s cloud
  quantum computer.
\newblock \emph{Science China Physics, Mechanics \& Astronomy}, 61\penalty0
  (7):\penalty0 1--10, 2018.
\newblock \doi{10.1007/s11433-017-9181-9}.

\bibitem[Napolitano et~al.(2011)Napolitano, Koschorreck, Dubost, Behbood,
  Sewell, and Mitchell]{napolitano2011interaction}
Mario Napolitano, Marco Koschorreck, Brice Dubost, Naeimeh Behbood, RJ~Sewell,
  and Morgan~W Mitchell.
\newblock Interaction-based quantum metrology showing scaling beyond the
  heisenberg limit.
\newblock \emph{Nature}, 471\penalty0 (7339):\penalty0 486--489, 2011.
\newblock \doi{10.1038/nature09778}.

\bibitem[ref()]{ref:quafu}
Detail information about \textup{Q}uafu cloud platform can be found at
  \href{http://quafu.baqis.ac.cn/}{website},
  \href{https://github.com/ScQ-Cloud/pyquafu}{github}, and
  \href{https://scq-cloud.github.io/}{document}.

\bibitem[Du et~al.(2010)Du, Xu, Peng, Wang, Wu, and Lu]{du2010nmr}
Jiangfeng Du, Nanyang Xu, Xinhua Peng, Pengfei Wang, Sanfeng Wu, and Dawei Lu.
\newblock Nmr implementation of a molecular hydrogen quantum simulation with
  adiabatic state preparation.
\newblock \emph{Physical review letters}, 104\penalty0 (3):\penalty0 030502,
  2010.
\newblock \doi{10.1103/PhysRevLett.104.030502}.

\bibitem[Panju(2011)]{panju2011iterative}
Maysum Panju.
\newblock Iterative methods for computing eigenvalues and eigenvectors.
\newblock \emph{arXiv preprint arXiv:1105.1185}, 2011.
\newblock \doi{10.48550/arXiv.1105.1185}.

\end{thebibliography}


\end{document}